\documentclass[useAMS,usenatbib,babel]{mn2e}

\usepackage[english,english]{babel}
\usepackage{amsmath}
\usepackage{amssymb,amsfonts,textcomp}
\usepackage{array}
\usepackage{supertabular}
\usepackage{hhline}
\usepackage{hyperref}
\usepackage[usenames]{color}
\hypersetup{dvips, colorlinks=true, linkcolor=black, citecolor=black, filecolor=blue, urlcolor=blue}
\usepackage[dvips]{graphicx}

\def\gtrsim{\lower.5ex\hbox{$\; \buildrel > \over \sim \;$}}
\usepackage{graphicx}

\definecolor{grey}{rgb}{0.75,0.75,0.75}
\definecolor{Orange}{rgb}{1.0,0.5,0.15}
\definecolor{brown}{rgb}{0.7,0.25,0.0}
\definecolor{pink}{rgb}{1.0,0.5,0.5}
\definecolor{darkerred}{rgb}{0.8,0,0}
\definecolor{darkerblue}{rgb}{0,0,0.8}
\definecolor{Blue}{rgb}{0,0.08,0.65}
\definecolor{Red}{rgb}{0.65,0.08,0.05}
\definecolor{Green}{rgb}{0.15,0.45,0.25}

\begin{document}

\author[Y. Dubois et al. ]{
\parbox[t]{\textwidth}{
Yohan Dubois$^{1,2}$\thanks{E-mail: dubois@iap.fr}, Marta Volonteri$^{1,3}$ and Joseph Silk$^{1,2,4}$}
\vspace*{6pt} \\
$^{1}$ Institut d'Astrophysique de Paris, UMR 7095, CNRS, UPMC Univ. Paris VI, 98 bis boulevard Arago, 75014 Paris, France\\
$^{2}$ Sub-department of Astrophysics, University of Oxford, Keble Road, Oxford OX1 3RH\\
$^{3}$ Astronomy Department, University of Michigan, Ann Arbor, MI 48109, USA\\
$^{4}$ Department of Physics and Astronomy, The Johns Hopkins University Homewood Campus, Baltimore, MD 21218, USA\\
}
\date{Accepted 2014 February 25. Received 2013 December 23; in original form 2013 April 16}

\title[BH masses and spins in cosmological simulations]
{Black hole evolution: III. Statistical properties of mass growth and spin evolution using large-scale hydrodynamical cosmological simulations}

\maketitle

\begin{abstract}
{Supermassive black holes (BH) at the centres of galaxies can rapidly change their mass and spin by gas accretion and mergers. Using hydrodynamical cosmological simulations, with prescriptions for BH growth and feedback from Active Galactic Nuclei, we study how the evolution of BH mass growth is driven by gas accretion and mergers. Using a semi-analytical approach to evolve spins, we also highlight the mechanisms responsible for driving the magnitude and the direction of spins as a function of cosmic time. We find that in the high-redshift universe galaxies maintain large values of gas accretion onto BHs, which therefore is the main driver of their mass and spin evolution. Sustained accretion of cold gas at high-redshift tends to align BH spins with the angular momentum of the surrounding gas and maximise their magnitude. Conversely, at low redshift, as BHs get more massive and galaxies more gas-poor, the contribution from binary coalescences to the total BH mass growth increases, especially at the high-mass end, and tends to decrease the  magnitude of spins and change their direction. }
\end{abstract}


\section{Introduction}

In the currently favored scenario for the formation of cosmic structures in the Universe, present-day galaxies have been built up, via a series of mergers, from small building blocks that condensed out at early cosmic times. In this paradigm, galaxies experience multiple mergers during their lifetime. A single large galaxy, now containing a BH, can be traced back to the stage when it was split up in hundreds of components with masses a million times smaller than todayÕs galaxies. At the same time galaxies grow by accretion of intergalactic gas, mediated through filaments. The BHs that are now  commonly observed in the centres of galaxies have followed a similar growth path, that consists of both BH-BH coalescences and accretion of ambient gas. 

A time-dependent Soltan's argument can be used to argue that gas accretion, as opposed to mergers, drives most of the mass growth of black holes~\citep{yutremaine2002} either through direct accretion of cosmic filamentary gas or of star-forming clumps~\citep{bournaudetal11, dimatteoetal12, duboisetal12angmom, bellovaryetal13, fengetal13}. This mechanism converts a fraction of the accreted rest-mass energy into effective feedback for the host galaxy, which in turn can explain the observed scaling relation between black holes and galaxies~\citep{silk&rees98, king03, wyithe&loeb03}. Numerical implementations of the feedback of Active Galactic Nuclei (AGN) have demonstrated that the energy released by BHs is able to shape the galaxy luminosity function in the bright-end~\citep{crotonetal06, boweretal06}, and halt the cooling catastrophe operating in massive halos~\citep{sijackietal07, duboisetal10, teyssieretal11}. The amount of energy that is extracted from accretion events is closely related to the spin of the central BH the gas is accreting onto.
Maximally spinning BHs release more specific energy than non-rotating or counter-rotating BHs, which could have strong consequences for the self-regulated growth of BHs and the rate at which they accrete gas. 

Analytical or semi-analytical models based on pure dark matter simulations have tried to address the cosmological evolution of BH spins related to their host properties~\citep{volonterietal05, shapiro2005,volonterietal07, lagosetal09, fanidakisetal11, barausse12,volonterietal13}.
Very few hydrodynamical simulations of the spin evolution of BHs embedded into their host galaxies have been performed: they are either performed in isolated contexts though with extremely high resolution~\citep{maioetal13} or neglect the role of gas accretion for spin evolution~\citep{sijackietal09}.
Our aim is to explore further the spin evolution paradigm by evolving spins with a semi-analytical approach (i.e., post-processing spins) that takes into account  gas accretion and BH coalescences, at the pace dictated by their cosmological environment, and simulated directly.

This work is part of a series of paper where we explore the interplay between supernova and AGN feedback in growing galaxies and BHs (paper I, Dubois et al., in prep.), the effect of supernova-driven turbulence and cosmological perturbations (satellites, cold flows) on BH spin evolution (paper II,~\citealp{duboisetalpaper3}).
In this paper (paper III), we focus on statistical properties of BH mass growth and spins using large-scale hydrodynamical cosmological simulations of galaxy evolution with BH growth and AGN feedback, i.e. where the properties of galaxies and the BH masses are treated self-consistently.  BH spins are instead evolved  in a semi-analytic (post-processed) manner tracking accretion of gas (tracking its angular momentum) and BH coalescences.

In section~\ref{section:numerics}, we describe the initial conditions and the physics (gas cooling, star formation, stellar and AGN feedback) employed to follow the formation of galaxies in a $\Lambda$CDM universe.
In section~\ref{section:spinmodel}, we give the properties of our model for BH spin evolution by using properties extracted from hydrodynamical cosmological simulations.
In section~\ref{section:result}, we describe the predicted spin evolution from our model and how this quantity is driven by BH growth.
We finally summarise and discuss our main results in section~\ref{section:conclusion}.

\section{Numerical set-up}
\label{section:numerics}

\subsection{Initial conditions}

We assume a $\Lambda$CDM cosmology with total matter density $\Omega_{m}=0.272$, baryon density $\Omega_b=0.045$, dark energy density $\Omega_{\Lambda}=0.728$, amplitude of the matter power spectrum $\sigma_8=0.8$ and  Hubble constant $H_0=70.4\, \rm km\, s^{-1} \, \rm Mpc^{-1}$ consistent with the WMAP 7-year data \citep{komatsuetal11}.
The box size of our simulation is $L_{\rm box}=50\,  h^{-1}\, \rm Mpc$, with $256^3$ dark matter (DM) particles corresponding to a DM mass resolution of $M_{\rm res}=6.6\times 10^8 \, \rm M_\odot$.
An extra zoomed-in simulation of a massive halo at $z=6$ with 10 pc resolution is analysed further in Appendix~\ref{section:zoom-halo} to test the robustness of sampling the gas angular momentum on kpc scales, along with a simulation of an isolated disc galaxy at similar resolution.

The simulations are run with the Adaptive Mesh Refinement code {\sc ramses} \citep{teyssier02}.
The evolution of the gas is followed using a second-order unsplit Godunov scheme for the Euler equations.
The HLLC Riemann solver with a first-order MinMod Total Variation Diminishing scheme to reconstruct the interpolated variables from their cell-centered values is used to compute fluxes at cell interfaces.
Collisionless particles (DM, star and BH particles) are evolved using a particle-mesh solver with a Cloud-In-Cell interpolation.
The coarse mesh (with minimum level of refinement equals 8) is refined up to $\Delta x=2.2$~kpc (maximum level equals 15) using a quasi-Lagrangian strategy: when more than 8 DM particles lie in a cell, or if the baryon density is larger than 8 times the initial DM resolution.
The minimum cell size is kept roughly constant in physical size with redshift, i.e. an additional level of refinement is added every $a_{\rm exp}=n\times0.1$ (where $n=1,2,4,8$ and $a_{\rm exp}$ is the expansion factor of the universe) up to level 15 at $a_{\rm exp}=0.8$.

\subsection{Physics of galaxy formation}

Gas is allowed to cool by H and He cooling with a contribution from metals using a~\cite{sutherland&dopita93} model for temperatures above $T_0=10^4$~K, which is the minimum temperature of the gas allowed through radiative losses.
Heating from a uniform UV background takes place after redshift $z_{\rm reion}=10$ following~\cite{haardt&madau96}.
Metallicity is modelled as a passive variable for the gas advected with the flow (whose composition is assumed to be solar) and is altered by the injection of gas ejecta during supernovae explosions and stellar mass losses.
We assume a zero initial metallicity.
The gas follows an adiabatic equation of state for ideal monoatomic gas with adiabatic index $\gamma=5/3$.

The star formation process is modelled with a Schmidt law:
$\dot \rho_*= \epsilon_* {\rho / t_{\rm ff}}\, ,$ where $\dot \rho_*$ is the star formation rate density, $\epsilon_*$ the constant star formation efficiency, and $t_{\rm ff}$ the local free-fall time of the gas.
We choose a low star formation efficiency $\epsilon_*=0.02$ consistent with observations of giant molecular clouds~\citep{krumholz&tan07} and surface density relations of galaxies~\citep{kennicutt98}.
Star formation is allowed in regions exceeding a gas density threshold of $\rho_0=0.1\, \rm H\, cm^{-3}$.
The gas pressure is artificially enhanced above $\rho > \rho_0$ assuming a polytropic equation of state $T=T_0(\rho/\rho_0)^{\kappa-1}$ with polytropic index $\kappa=4/3$ to avoid excessive gas fragmentation.
Feedback from stars is taken into account assuming a Salpeter initial mass function with a low-mass (high-mass) cut-off of $0.1\, \rm M_{\odot}$ ($100 \, \rm M_{\odot}$), as described in detail in Kimm et al. (in prep.). 
Specifically, the mechanical energy from supernovae type II and stellar winds is taken from {\sc starburst99}~\citep{leithereretal99, leithereretal10}, and the frequency of supernovae type Ia explosions is computed following~\cite{greggio&renzini83}. 
The energy from SNe and stars is coupled to the gas with a kinetic implementation from~\cite{dubois&teyssier08winds} until the age of stars reaches 50 Myr, where mass, momentum and kinetic energy profiles are imposed to mimic a Sedov blast wave. 
After 50 Myr, the energy is released thermally into the gas where the star is present.

\subsection{Model for BH growth and AGN feedback}

We use the same ``canonical'' AGN feedback modelling employed in~\cite{duboisetal12agnmodel}.
BHs are created at loci where the gas density is larger than the density threshold for star formation $\rho_0$ with an initial seed mass of $10^5\, \rm M_\odot$.
In order to avoid the formation of multiple BHs in the same galaxy, BHs are not allowed to form at distances smaller than 50 kpc from any other BH particle.
The accretion rate onto BHs follows the Bondi-Hoyle-Lyttleton~\citep{bondi52} rate
$\dot M_{\rm BH}=4\pi \alpha G^2 M_{\rm BH}^2 \bar \rho / (\bar c_s^2+\bar u^2) ^{3/2},$
where $M_{\rm BH}$ is the BH mass, $\bar \rho$ is the average gas density, $\bar c_s$ is the average sound speed, $\bar u$ is the average gas velocity relative to the BH velocity, and $\alpha$ is a dimensionless boost factor with $\alpha=(\rho/\rho_0)^2$ when $\rho>\rho_0$ and $\alpha=1$ otherwise~\citep{booth&schaye09} in order to account for our inability to capture the colder and higher density regions of the ISM.
The effective accretion rate onto BHs is capped at the Eddington accretion rate:
$\dot M_{\rm Edd}=4\pi G M_{\rm BH}m_{\rm p} / (\epsilon_{\rm r} \sigma_{\rm T} c),$
where $\sigma_{\rm T}$ is the Thompson cross-section, $c$ is the speed of light, $m_{\rm p}$ is the proton mass, and $\epsilon_{\rm r}$ is the radiative efficiency, assumed to be equal to $\epsilon_{\rm r}=0.1$ for the \cite{shakura&sunyaev73} accretion onto a Schwarzschild BH.
Note that the radiative efficiency of accretion is a function of the spin parameter of the BH.
Here, we assume that value to be fixed as the BH spins are not followed on-the-fly, but are a post-processed quantity from cosmological runs.

In order to avoid spurious oscillations of the BH in the gravitational potential well due to external perturbations and finite resolution effects, we introduce a drag force that mimics the dynamical friction exerted by the gas onto a massive particle.
This dynamical friction is proportional to $F_{\rm DF}=f_{\rm gas} 4 \pi \alpha \rho (G M_{\rm BH}/\bar c_s)^2$, where $f_{\rm gas}$ is a fudge factor whose value is between 0 and 2 and is a function of the mach number ${\mathcal M}=\bar u/\bar c_s<1$~\citep{ostriker99, chaponetal13}, and where we introduce the boost factor $\alpha$ for the same reasons than stated above.

The AGN feedback is a combination of two different modes, the so-called \emph{radio} mode operating when $\chi=\dot M_{\rm BH}/\dot M_{\rm Edd}< 0.01$ and the \emph{quasar} mode active otherwise.
The quasar mode corresponds to an isotropic injection of thermal energy into the gas within a sphere of radius $\Delta x$, at an energy deposition rate: $\dot E_{\rm AGN}=\epsilon_{\rm f} \epsilon_{\rm r} \dot M_{\rm BH}c^2$,
where $\epsilon_{\rm f}=0.15$ for the quasar mode is a free parameter chosen to reproduce the $M_{\rm BH}$-$M_{\rm b}$, $M_{\rm BH}$-$\sigma_{\rm b}$, and BH density in our local Universe (see \citealp{duboisetal12agnmodel}).
At low accretion rates on the other hand, the radio mode deposits the AGN feedback energy into a bipolar outflow with a jet velocity of $10^4\,\rm km.s^{-1}$ into a cylinder with a cross-section of radius $\Delta x$ and height $2 \, \Delta x$ following~\cite{ommaetal04} (more details about the jet implementation are given in~\citealp{duboisetal10}).
The efficiency of the radio mode is larger with $\epsilon_{\rm f}=1$.

\section{Model of BH spin evolution}
\label{section:spinmodel}

\subsection{Gas accretion}

BH spins are allowed to be change their magnitude through accretion of gas through the following expression~\citep{bardeen70}:
\begin{equation}
\label{aspinup}
a^{\rm n+1}={1\over 3} {r_{\rm isco}^{1/2}\over M_{\rm ratio}}\left [ 1- \left( 3 {r_{\rm isco}\over M_{\rm ratio}^2}-2 \right )^{1/2}\right ]\, ,
\end{equation}
where $M_{\rm ratio}=M^{\rm n+1}_{\rm BH}/M^{\rm n}_{\rm BH}$, the n superscript stands for the values at time $t_{\rm n}$, and $R_{\rm isco}$ is the radius of the innermost stable circular orbit (ISCO) defined as (in reduced units):
\begin{equation}
r_{\rm isco}=R_{\rm isco}/R_{\rm g}=3+Z_{2}\pm [(3-Z_1)(3+Z1+2Z_2)]^{1/2}\, , 
\end{equation}
where the gravitational radius $R_{\rm g}$ is defined as half of the Schwarzschild radius of the BH, $R_{\rm BH}$, and $Z_1$ and $Z_2$ are:
\begin{eqnarray}
Z_1&=&1+(1-a^2)^{1/3}[(1+a)^{1/3}+(1-a)^{1/3}] \, , \\
Z_2&=&(3a^2+Z_1^2)^{1/2}\, .
\end{eqnarray}
For positive spins, co-rotating with their accretion disc, $1\le r_{\rm isco}<6$, while for negative spins, counter-rotating with their accretion disc, $6<r_{\rm isco}\le 9$, and $r_{\rm isco}=6$ for non-spinning BHs.

The equation that governs the evolution of BH spin through direct accretion of gas is correct when BH spin and disc angular momentum are perfectly aligned (or anti-aligned), but in the most-general case, misalignment occurs.
In the misaligned case, the accretion disc experiences a torque due to the Lense-Thirring effect that causes the accretion disc to precess about the spin axis of the BH.
For large enough viscosities the innermost parts of the disc are forced to rotate within the equatorial plane of the disc and a warped disc is created.
We can define the total angular momentum of the system $\{$disc+BH$\}$ as $\bmath{J}_{\rm tot}=\bmath{J}_{\rm d}+\bmath{J}_{\rm BH}$.
The values of the angle $\theta$ between $\bmath{J}_{\rm BH}$ and $\bmath{J}_{\rm d}$ are between $-1\le \cos \theta\le1$ where the two-extrema correspond to anti-aligned and aligned cases respectively. 

The result of the Lense-Thirring precession is that the BH and disc angular momentum ends up being aligned or anti-aligned with the total angular momentum.
The case for anti-alignment of the BH with the disc requires that~\citep{kingetal05}
\begin{equation}
\cos \theta < - {J_{\rm d}\over 2 J_{\rm BH}}\, , 
\end{equation}
thus, for $\cos \theta\ge 0$, they always align, while for $\cos \theta < 0$, they eventually anti-align if the ratio $J_{\rm d}/J_{\rm BH}$ is sufficiently small compared to $\cos \theta$.

One difficulty we find is in accessing  the magnitude of $\bmath{J}_{\rm d}$ as we cannot resolve the accretion disc.  We therefore assume the tilted solution for a  thin accretion disc~\citep{shakura&sunyaev73,Scheuer1996,NatarajanPringle1998,Perego2009}, characterized by a viscosity $\nu_1=\alpha_t c_s^2 /(GM_{\rm BH}/r^3)^{1/2}$, where $\alpha_t$ is a parameter $<1$, and $c_s$ is the sound speed.  In a misaligned disc under the effect of  Lense-Thirring precession a  natural scale is the warp radius, that marks the transition between an equatorial inner disc and a misaligned outer disc: the direction of the angular momentum of the inflowing material changes direction as it passes through the warp. Only material within the warp radius can effectively transfer its angular momentum to the BH~\citep{volonterietal07}. This is the relevant scale to estimate whether alignment or anti-alignment occur. 
In a~\cite{shakura&sunyaev73}  thin accretion disc, one can write the ratio of the warp to Schwarzschild radius as: 
\begin{equation}
{R_{\rm warp}\over R_{\rm BH}}\simeq 6.4 \times 10^3 a^{5/8} M_{\rm BH, 8}^{1/8} \left( \epsilon_{\rm r,01} \over \chi \right)^{1/4} \left ({\nu_2\over \nu_1}\right )^{-5/8} \alpha_{\rm t,01}^{-1/2}\, , 
\end{equation}
where $\epsilon_{\rm r,01}$ is the radiative efficiency normalised to $0.1$, $\nu_1$ and $\nu_2$ are the kinematic viscosities horizontal and perpendicular to the equatorial plane of the disc: $\nu_2$ is the viscosity responsible the warp propagation, while $\nu_1$ is the viscosity responsible for driving accretion and transferring angular momentum. 
We choose a value for $\alpha_{\rm t,01}\equiv\alpha_{\rm t}/0.1=1$~\citep{kingetal07}, and the ratio of $\nu_2/\nu_1=2(1+7\alpha_{\rm t})/(4+\alpha_{\rm t}^2)/\alpha_{\rm t}^2$ is of the form given by~\cite{ogilvie99} (equation 145). 
With these choices, $(\nu_2/\nu_1)\sim 85$  (note that for small $\alpha_t$,  $ \nu_2\sim \nu_1/\alpha_t^2$) and: 
\begin{equation}
\label{rwarp}
{R_{\rm warp}\over R_{\rm BH}}\simeq 4 \times 10^2 a^{5/8} M_{\rm BH, 8}^{1/8} \left( \epsilon_{\rm r,01} \over \chi \right)^{1/4} \left ({\nu_2/\nu_1\over 85}\right )^{-5/8} \alpha_{\rm t,01}^{-1/2}  \,. 
\end{equation}

The BH angular momentum is simply $J_{\rm BH} = a M_{\rm BH}^{3/2} R_{\rm BH}^{1/2}$, and the disc angular momentum can now be expressed as $J_{\rm d} \sim M_{\rm d}(R_{\rm warp}) M_{\rm BH} ^{1/2} R_{\rm warp}^{1/2}$. The disc mass within $R_{\rm warp}$ is $M_{\rm d}(R_{\rm warp})=\dot M t_{\nu_1}(R_{\rm warp})$ where $t_{\nu_1}$ is the viscous timescale for radial propagation:
\begin{equation}
{t_{\nu_1}}=5.3\times 10^5 a^{7/8} M_{\rm BH, 8}^{11/8} \left( \epsilon_{\rm r,01} \over \chi \right)^{3/4} \left ({\nu_2\over \nu_1}\right )^{-7/8} \alpha_{\rm t}^{-3/2} {\rm yr}\, 
\end{equation}
$$\sim 3.4 \times 10^5 a^{7/8} M_{\rm BH, 8}^{11/8}  \left( \epsilon_{\rm r,01} \over \chi \right)^{3/4} \left ({\nu_2/\nu_1\over 85}\right )^{-7/8} \alpha_{\rm t,01}^{-3/2} {\rm yr}.$$

The ratio of disc to BH angular momentum can, then, be written as:
\begin{equation}
\label{jdjbhratio}
{J_{\rm d}\over2 J_{\rm BH}}\simeq {M_{\rm d}(R_{\rm warp}) \over  a M_{\rm BH}} \left ({R_{\rm warp}\over R_{\rm BH}}\right )^{1/2}\,
\end{equation}
$$ \sim 6.8\times 10^{-2}  \left( \chi \over \epsilon_{\rm r,01} \right)^{1/8} M_{\rm BH, 8}^{23/16} a^{3/16} \alpha_{\rm t,01}^{-7/4}\left ({\nu_2/\nu_1\over 85}\right )^{-19/16}.$$
Equation~(\ref{jdjbhratio}) allows one to evaluate if anti-alignment occurs, and provides a value for the magnitude of $\bmath{J}_{\rm d}$.
We refer the reader to \cite{dottietal13} for a detailed discussion of how the ratio $J_{\rm d}/ J_{\rm BH}$ determines the behaviour of alignment and spin evolution. 

For the direction of $\bmath{J}_{\rm d}$, we make the assumption that the accretion disc is aligned with the gas angular momentum, $\bmath{J}_{\rm g}$, measured in the surroundings of the BHs as extracted directly in the simulation (but their magnitude is not equal). Note that it is potentially a strong assumption because, if the gas is sufficiently turbulent, the inner dense cold (unresolved) molecular gas  surrounding the BH can misalign with the large-scale galactic disc. In Appendices A and B we provide additional details and validate our technique and assumption using high-resolution simulations that purposely study the degree of  consistency between the angular momentum on large scales and small scales, as well as the dependability of this assumption (see paper II for a more thorough analysis).

Thus, we  update the orientation of the BH  angular momentum after one accretion event (in order to avoid overestimating $J_{\rm d}$ we divide each simulation coarse time step, $\sim$ 2 Myr, in a sequence of $\Delta t=t_{\nu_1}$) with  $\bmath{J}_{\rm BH}^{\rm n+1}= \bmath{J}_{\rm tot}= \bmath{J}_{\rm BH}^{\rm n} + J_{\rm d}^{\rm n} \bmath{j}_{\rm g}^{\rm n}$ for the alignment case, where the small $\bmath{j}$ stands for angular momentum unit vectors, and where $J_{\rm d}^{\rm n}$ is provided by equation (\ref{jdjbhratio}).
If the criterion for anti-alignment is met, $\bmath{J}_{\rm BH}^{\rm n+1}$ becomes $-\bmath{J}_{\rm tot}$ if $\bmath{J}_{\rm tot}.\bmath{J}_{\rm BH}^{\rm n}>0$, this occurs when the total angular momentum dominated by the BH  \citep[case b in Fig.~1 in][] {kingetal05}. Otherwise, if the anti-alignment condition is satisfied and $\bmath{J}_{\rm tot}.\bmath{J}_{\rm BH}^{\rm n}<0$, then the total angular momentum is dominated by the disc and it is anti-aligned with respect to that of the BH, therefore  $\bmath{J}_{\rm BH}^{\rm n+1}=\bmath{J}_{\rm tot}$ \citep[case d in Fig.~1 in][]{kingetal05}. 

For sufficiently large accretion rates, accretion discs can become unstable against their own gravity and  fragment into gas clumps~\citep[e.g.][]{kolykhalov&sunyaev80, pringle81,goodman&tan04}. The criterion for stability given by the Toomre parameter $Q \sim c_{\rm s}\Omega/(\pi G\Sigma)$, where $\Omega$ is the Keplerian velocity and $\Sigma$ the gas surface density, provides the radius $R_{\rm sg}$ below which the accretion disc is stable. For the standard $\alpha$-disc, external region:
\begin{equation}
\label{rsg}
{R_{\rm sg}\over R_{\rm BH}}\simeq 5 \times 10^2 \alpha_{\rm t, 01}^{28/45} M_{\rm BH, 8}^{-52/45} \left( \epsilon_{\rm r,01} \over \chi \right)^{22/45} \, , 
\end{equation}
and the mass stable against fragmentation is~\citep[see][]{dottietal13}
\begin{equation}
\label{msg}
M_{\rm sg} \simeq 6 \times 10^5 \alpha_{\rm t,01}^{-1/45} M_{\rm BH, 8}^{34/45} \left( \chi \over \epsilon_{\rm r,01} \right)^{4/45} \, \rm M_\odot\, .
\end{equation}
We include the effects of self-gravity when $R_{\rm sg}< R_{\rm warp}$. In this case, the outer region of the disc is subject to fragmentation, while the central region as a whole aligns (or anti-aligns) in the equatorial plane of the BH.  Then, the angular momentum accreted onto the BH is the form of equation~(\ref{jdjbhratio}) with $R_{\rm warp}$ ($M_{\rm d}(R_{\rm warp})$) replaced by $R_{\rm sg}$ ($M_{\rm sg}$). Thus, we repeat the same steps as for the non self-gravitating case for the alignment (anti-alignment) criterion with the angular momentum calculated with $R_{\rm sg} $ and $M_{\rm sg} $.
In contrast to \cite{kingetal08}, who assume that  self-gravity causes chaos in the angular momentum distribution, we argue that the accreted gas conserves the angular momentum direction it possessed when entering the accretion flow, which is the direction measured from the simulation by definition.  \cite{kingetal08} state that the gas which is initially at radii $R < R_{\rm sg}$ forms a standard accretion disc, while most of the gas initially at radii $R > R_{\rm sg}$  is either turned into stars, or expelled by those stars which do form, causing BHs to be fuelled by well separated episodes. In this situation, the disc should be treated as a (truncated) standard (warped) thin disc during a single accretion episode of mass $M_{\rm sg}$ (or $M_{\rm d}(R_{\rm warp})$). The next accretion episode should not retain memory of the previous event, and therefore be treated as a new separate event, where the new angular momentum direction is determined by the boundary conditions of how the galaxy feeds the BH, as we do.  As a final note, the effects of self-gravity become important when  $R_{\rm sg}< R_{\rm warp}$, and through Eq.~(\ref{rwarp}) and~(\ref{rsg}), the critical BH mass above which the BH cannot stabilise anymore the disc against fragmentation is
\begin{equation}
\label{rwrsg2}
M_{\rm BH, 8} > 1.15 \, a^{- {225\over 461}} \left ( {\epsilon_{\rm r,01} \over \chi} \right )^{86 \over 461} \left ({\nu_2/\nu_1\over 85}\right )^{225 \over 461}   \alpha_{\rm t, 01}^{404 \over 461}\, .
\end{equation}
Similar expressions hold when using the middle region solutions for the $\alpha$-disc. In this case ${R_{\rm sg}\over R_{\rm BH}}\simeq 10^3 \alpha_{\rm t, 01}^{14/27} M_{\rm BH, 8}^{-26/27} \left( \epsilon_{\rm r,01} \over \chi \right)^{8/27}$ and the radius where self-gravity truncates the disc becomes smaller than the warp radius when $M_{\rm BH, 8} > 2.23 \, a^{- {135\over 269}} \left ( {\epsilon_{\rm r,01} \over \chi} \right )^{10 \over 269} \left ({\nu_2/\nu_1\over 85}\right )^{135 \over 269}   \alpha_{\rm t, 01}^{220 \over 269}\, .$

We note that in our simulations, including the high-resolution runs presented in the appendix, we often do not see the changes of angular momentum direction relative to the disk on kiloparsec scales highlighted by \cite{levineetal10}. Such fluctuations are present in our cosmological run (see Fig. A1), and may be characteristic of complex high-redshift galaxies with very high gas-fractions. The investigation of this issue using high-resolution cosmological simulations, where spin evolution is self-consistently tracked on-the-fly rather than in post-processing, is the focus of paper II. The main result is that even though gas in high-redshift galaxies has a high level of turbulence ($\sim 50 \, \rm km\, s^{-1}$) driven by supernovae explosions, it rarely overwhelms the large-scale rotation of the gas in the galaxy. When it does, it is either during supernovae-driven outflow phases where gas accretion onto the BH is quenched (and spin does not change much), or during mergers that rapidly reorientate the whole galactic AM, which one we can resolve with kpc resolution. In consequence, the gas AM momentum measured at kiloparsec scale is a good approximation of the AM of the accreted gas onto the BH at a few parsec scales. 

\begin{figure*}
  \centering{\resizebox*{!}{4.2cm}{\includegraphics{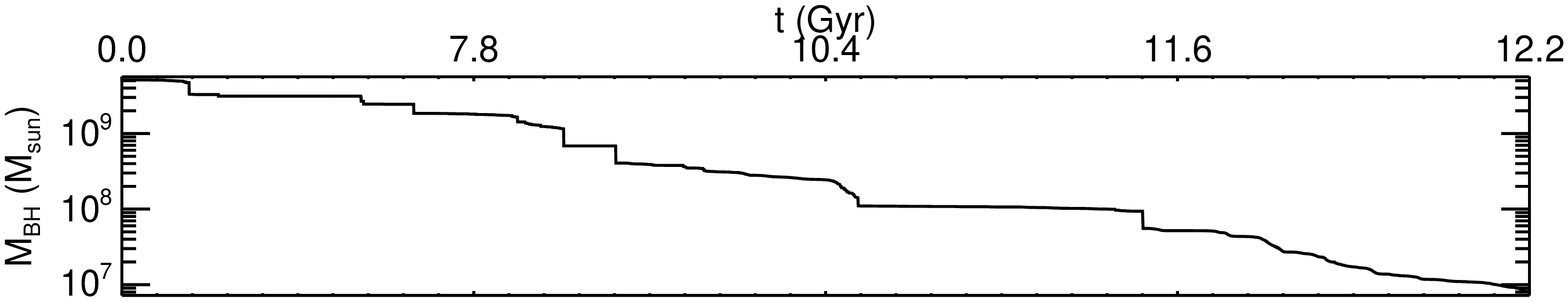}}}\vspace{-1.85cm}
  \centering{\resizebox*{!}{4.2cm}{\includegraphics{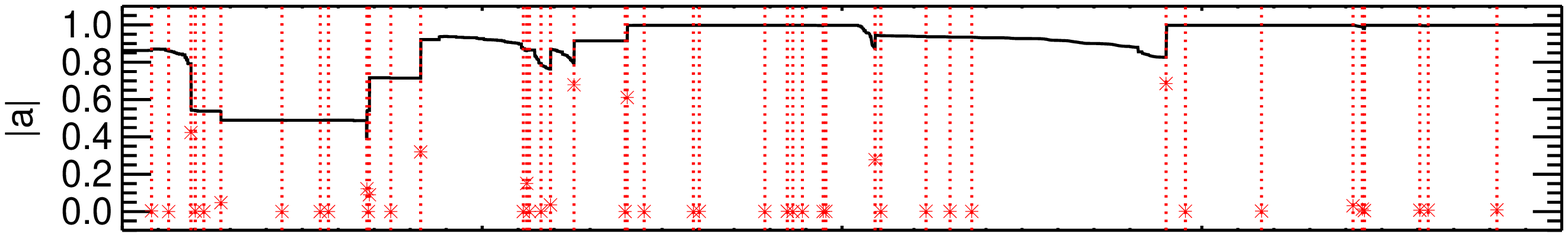}}}\vspace{-1.85cm}
  \centering{\resizebox*{!}{4.2cm}{\includegraphics{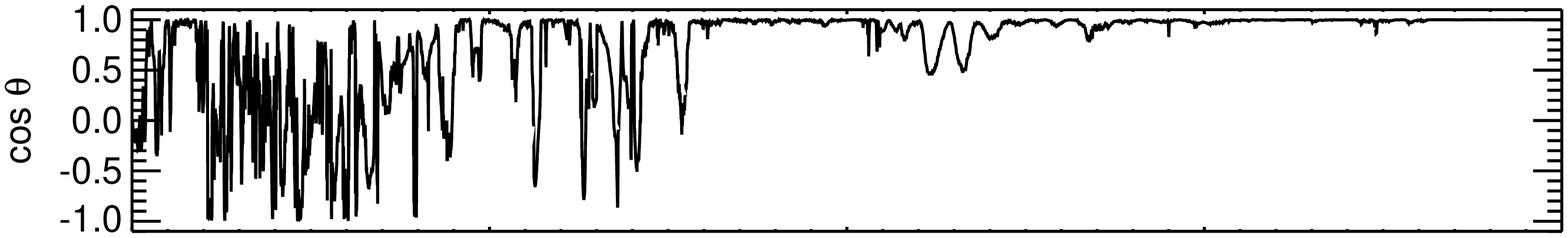}}}\vspace{-1.85cm}
  \centering{\resizebox*{!}{4.2cm}{\includegraphics{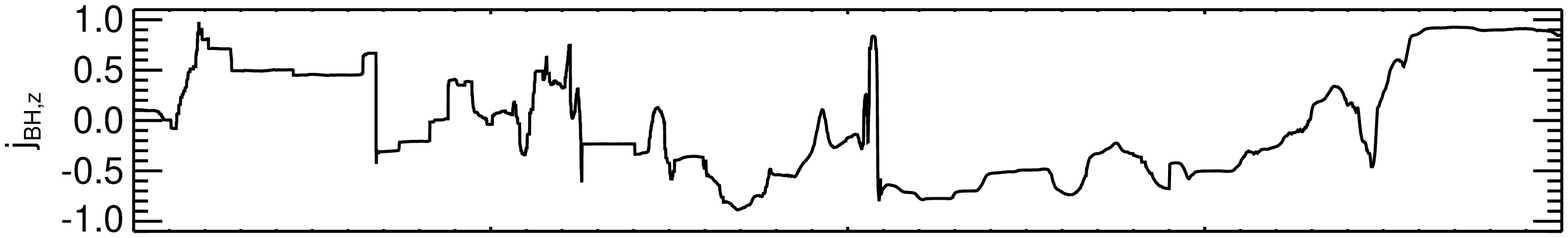}}}\vspace{-1.85cm}
  \centering{\resizebox*{!}{4.2cm}{\includegraphics{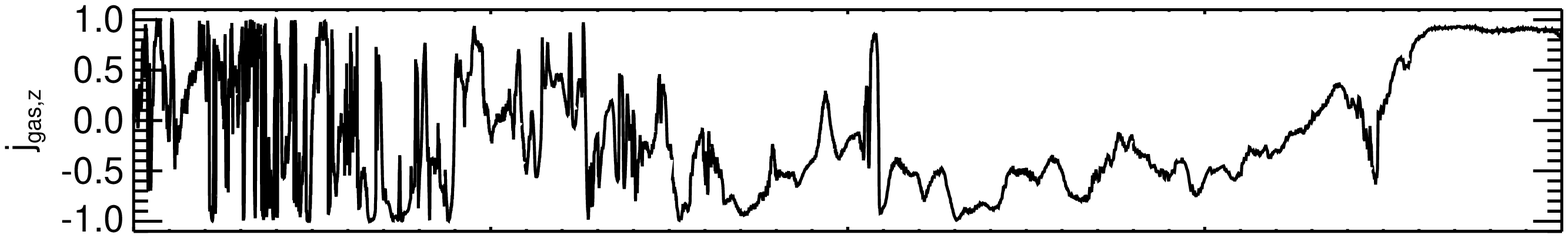}}}\vspace{-1.85cm}
 \centering{\resizebox*{!}{4.2cm}{\includegraphics{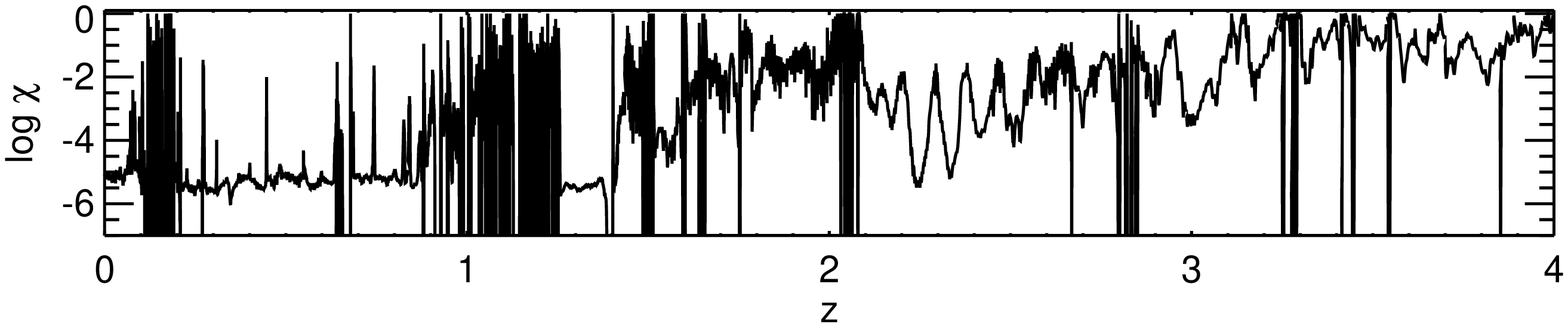}}}
  \caption{Example of the evolution of different properties of a BH of mass $5\times 10^9\, \rm M_\odot$ and which mass has been built-up by 73 per cent through mergers at $z=0$. \emph{From top to bottom}: BH mass evolution, BH spin evolution (red points indicate coalescence with its BH mass ratio), cosine of the angle between BH spin and gas angular momentum $\theta$, z-cartesian component of the normalised BH spin vector, z-cartesian component of the normalised gas angular momentum (both calculated with respect to the fixed reference frame of the simulation box), accretion rate over Eddington $\chi$. 
  }
    \label{fig:BHexample}
\end{figure*}

\begin{figure*}
  \centering{\resizebox*{!}{4.2cm}{\includegraphics{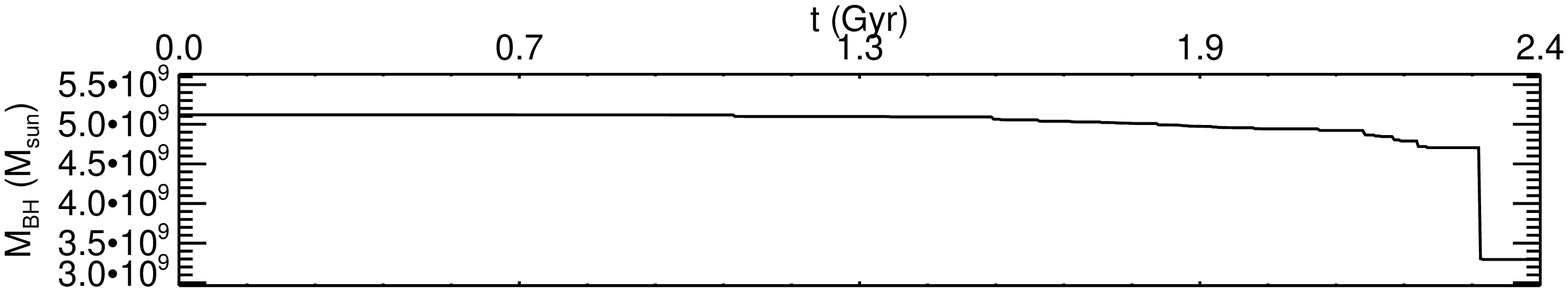}}}\vspace{-1.85cm}
  \centering{\resizebox*{!}{4.2cm}{\includegraphics{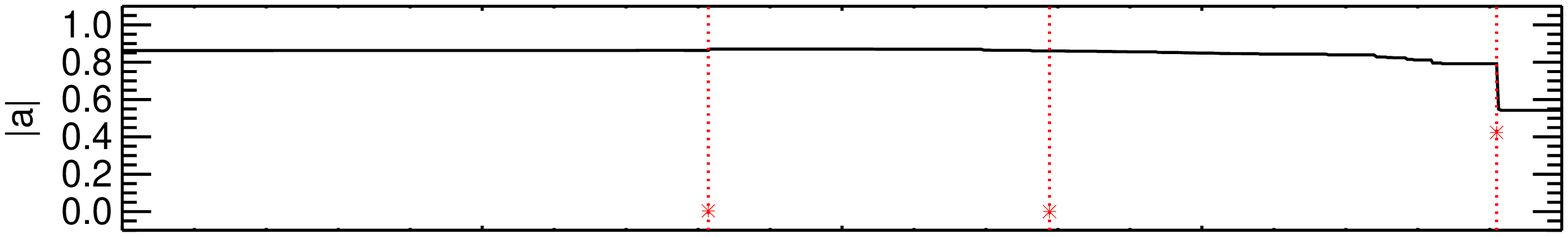}}}\vspace{-1.85cm}
  \centering{\resizebox*{!}{4.2cm}{\includegraphics{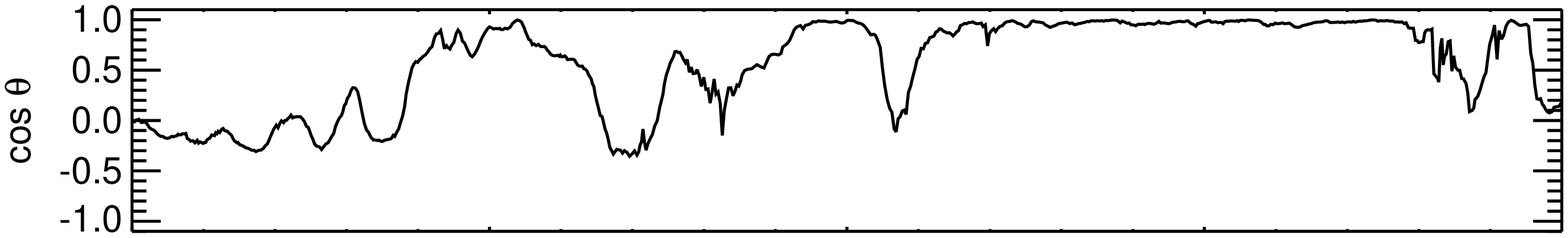}}}\vspace{-1.85cm}
  \centering{\resizebox*{!}{4.2cm}{\includegraphics{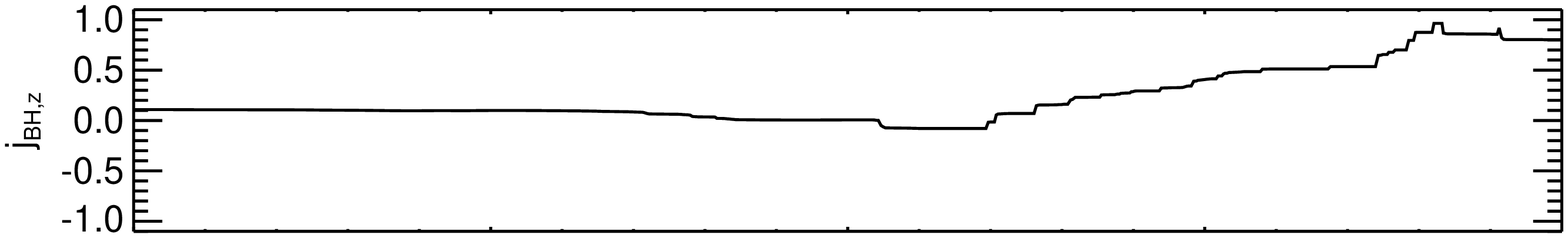}}}\vspace{-1.85cm}
  \centering{\resizebox*{!}{4.2cm}{\includegraphics{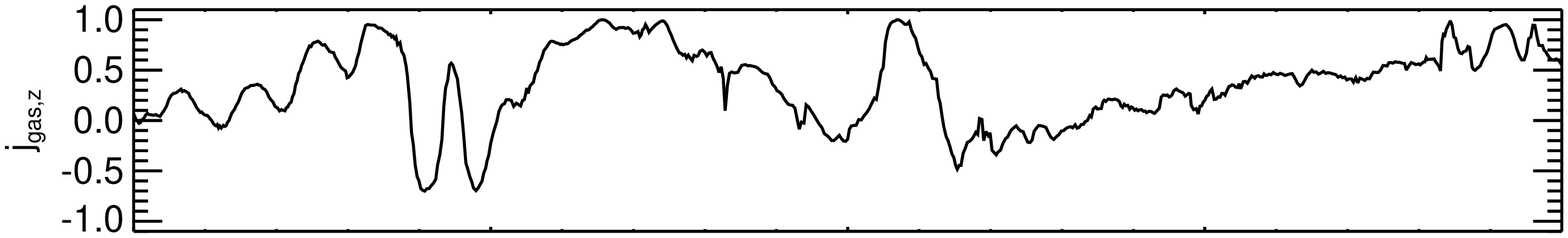}}}\vspace{-1.85cm}
 \centering{\resizebox*{!}{4.2cm}{\includegraphics{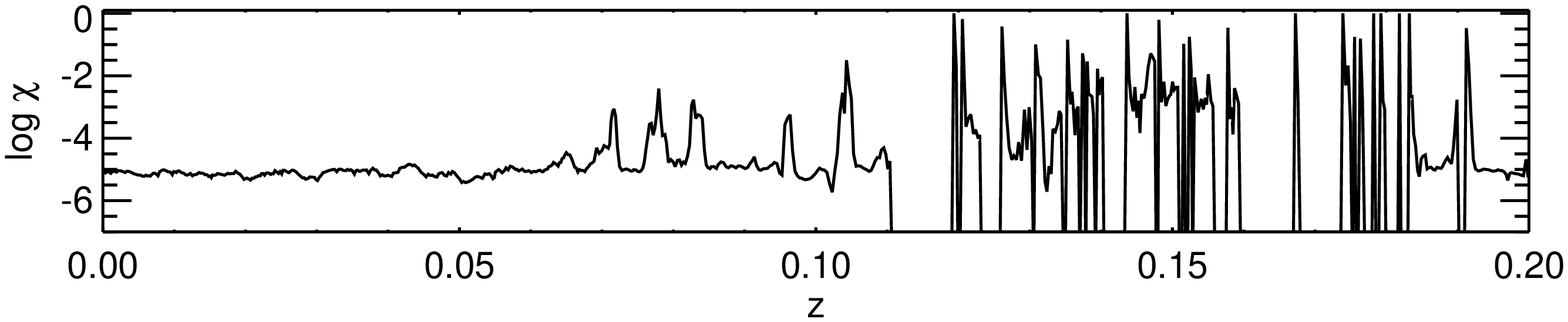}}}
  \caption{Same as Fig.~\ref{fig:BHexample} zoomed on the redshift range $z=[0,0.2]$. }
    \label{fig:BHexample_z0-0.2}
\end{figure*}

\begin{figure*}
  \centering{\resizebox*{!}{4.2cm}{\includegraphics{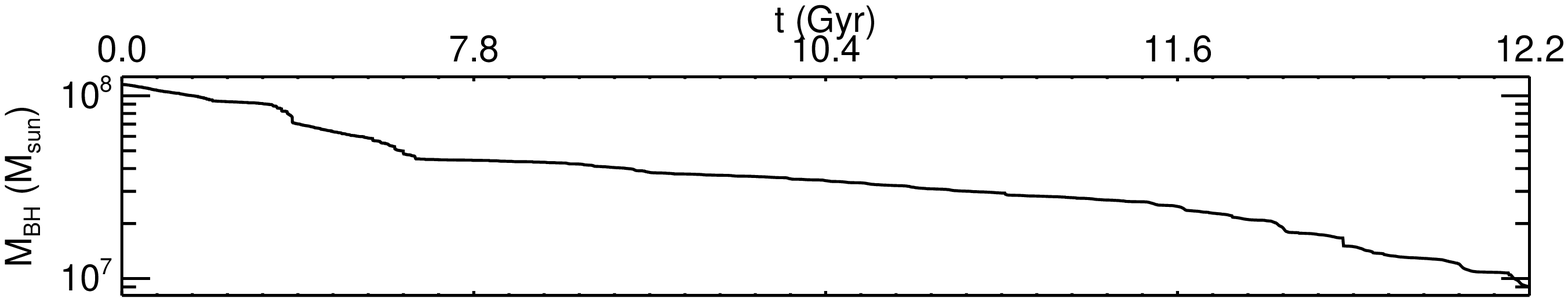}}}\vspace{-1.85cm}
  \centering{\resizebox*{!}{4.2cm}{\includegraphics{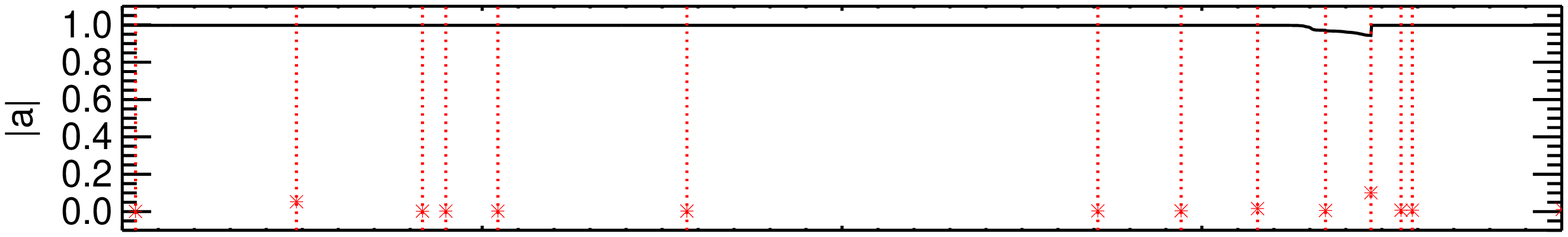}}}\vspace{-1.85cm}
  \centering{\resizebox*{!}{4.2cm}{\includegraphics{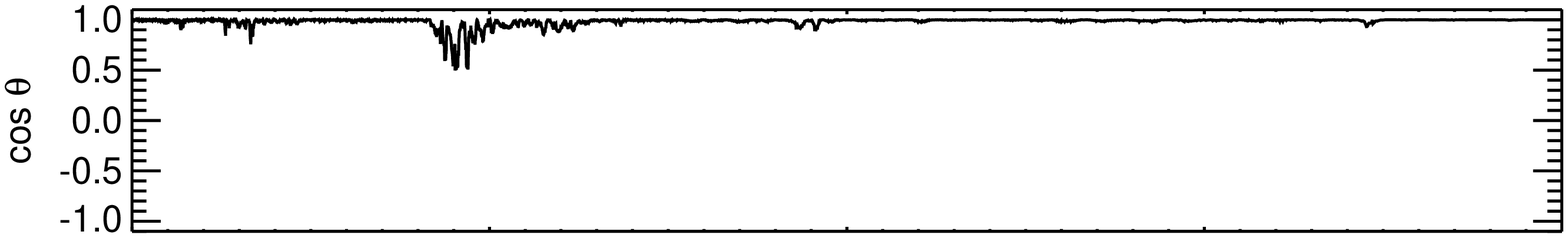}}}\vspace{-1.85cm}
  \centering{\resizebox*{!}{4.2cm}{\includegraphics{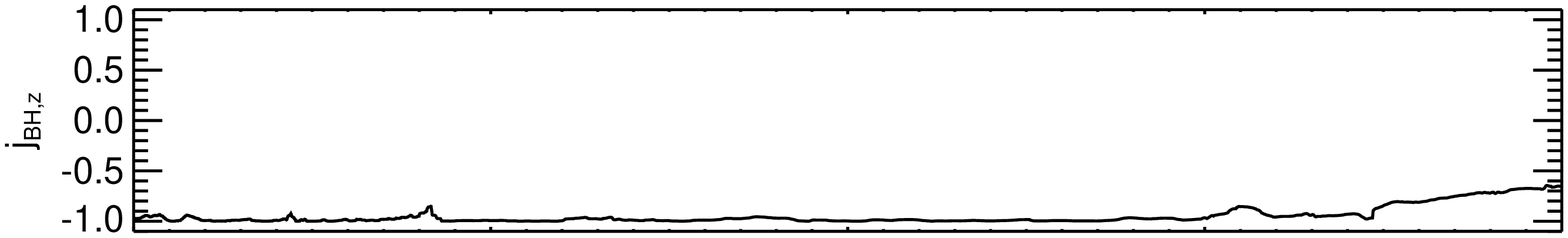}}}\vspace{-1.85cm}
  \centering{\resizebox*{!}{4.2cm}{\includegraphics{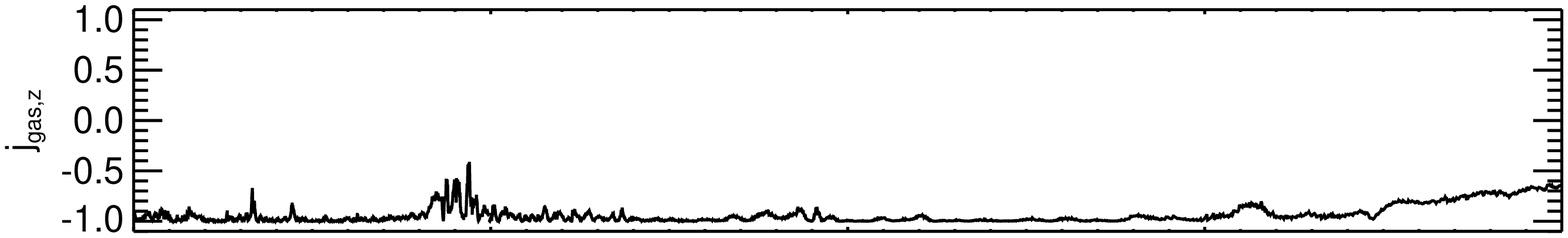}}}\vspace{-1.85cm}
 \centering{\resizebox*{!}{4.2cm}{\includegraphics{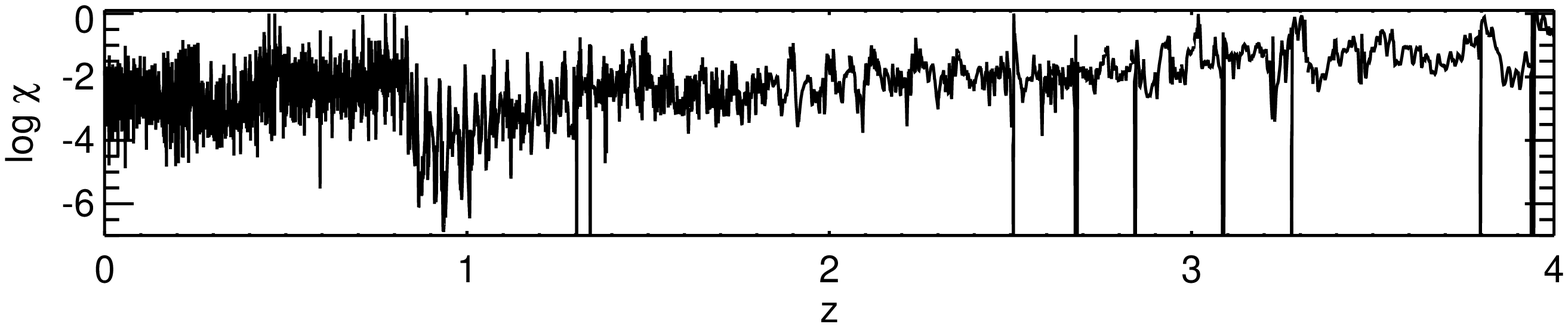}}}
  \caption{Same as Fig.~\ref{fig:BHexample} for a BH of mass $10^8\, \rm M_\odot$ and which mass has been built-up by 6 per cent through mergers at $z=0$.}
    \label{fig:BHexample2}
\end{figure*}

\subsection{Coalescence of BHs}

We track the change of the spin during the coalescence of two BHs using the analytical fit of~\cite{rezzollaetal08} derived from relativistic numerical simulations of BH binaries.
The final value of spin of the BH remnant is
\begin{equation}
\bmath{a}={1 \over (1+q)^2}(\bmath{a}_1+\bmath{a}_2q^2+ \bmath{\ell} q)\, , 
\end{equation}
where $\bmath{a}_1$ is the spin vector of the most massive BH progenitor, $\bmath{a}_2$ is the spin vector of the least massive BH progenitor, $q=M_2/M_1\le1$ the mass ratio of the binary, and $\bmath{\ell}=\bmath{l} /(M_1M_2)$, with $\bmath{l}$ the difference between the orbital  angular momentum $\bmath{L}$ when the two BHs are widely separated and the  angular momentum $\bmath{J}_{\rm gw}$ extracted from gravitational waves before the final coalescence $\bmath{l}=\bmath{L}-\bmath{J}_{\rm gw}$.

\cite{rezzollaetal08} provides a simple analytical expression for the norm of $\bmath{\ell}$
\begin{eqnarray}
\label{ell}
\ell&=&{s_4\over (1+q^2)^2} (a_1^2+a_2^2q^4+2 \bmath{a_1}.\bmath{a_2}q^2)\nonumber \\ 
&&+{s_5\mu +t_0+2\over 1+q^2}(a_1 \cos \phi_1+ a_2 q^2 \cos \phi_2)\nonumber \\ 
&&+2\sqrt3+t_2\mu+t_3\mu^2
\, , 
\end{eqnarray}
where $\phi_1$ ($\phi_2$) is the angle between $\bmath{a}_1$ ($\bmath{a}_2$) and $\bmath{\ell}$, and $\mu=q/(1+q)^2$.
We made the same assumption than~\cite{rezzollaetal08} to get the direction of $\bmath{\ell}$ by imposing collinearity with the orbital angular momentum $\bmath{L}=\bmath{L}_1+\bmath{L}_2$ before coalescence, which is a vector directly measured in our simulations.
To get access to $\bmath{L}$, we compute the centre of mass of the binary $\bmath{r}_{\rm com}$ and we evaluate the angular momentum of the two BHs relative to the centre of mass with $\bmath{L}_1=M_1 (\bmath{r}_1-\bmath{r}_{\rm com}) \times (\bmath{v}_1-\bmath{v}_{\rm com})$.
The numerical constants involved in equation~(\ref{ell}) are equal to $s_4=-0.129$, $s_5=-0.384$, $t_0=-2.686$, $t_2=-3.454$, and $t_3=2.353$.

\subsection{Updating spins}

The whole evolution of BH spins is post-processed over the numerical simulation results.
For this purpose, we print for each coarse time step (synchronisation of all levels at the minimum level of refinement) of the simulation the properties of BH particles that allows to derive the evolution of their spin.
We assume that the initial spin of BHs is zero at creation.
We further assume that the thin disc solution for spin evolution applies at any accretion rate over Eddington $\chi$.
It seems to be a fairly good approximation as BH spins mostly change their value by accretion during strong accretion events (of the order of $\chi \lesssim 1$).
The maximum spin value that we authorise is $a_{\rm max}=0.998$, due to the capture by the central BH of radiated photons emitted from the accretion disc~\citep{thorne74}.
Note that $a_{\rm max}=0.9$ is also possible for thick discs taking into accounts the magneto-hydrodynamical effects on the transfer of gas angular momentum~\citep{gammieetal04}, while~\cite{sadowskietal11} suggest $a_{\rm max}=0.9994$ if BH accretes gas at super-Eddington within a low-viscosity $\alpha_{\rm t}=0.01$ accretion disc.

We emphasize that the BH spin is a post-processed quantity from the hydrodynamical cosmological simulations.
For this reason, we neglect the effect of gravitational recoil on BH particles during the coalescence of BH binaries, which kicks the remnant BH depending on the magnitude  and configuration of their spins before coalescence~\citep{koppitzetal07}.
In future work, we will implement the scheme described above directly in the {\sc ramses} code, which will make possible to determine the efficiency of AGN feedback and the direction of jets from the spin magnitude and direction.

Note that a range of theoretical models assume that gas accretes onto BHs in random directions due to the capacity of the cold gas of the interstellar medium to drive turbulence~\citep{hopkinsetal12} or due to the formation of a self-gravitating accretion disc around BHs~\citep[e.g.][]{kingetal08}.
Therefore, we run extra models where some level of randomness in the direction of the accreted gas angular momentum is added explicitly to the scheme, either continuously or activated only during short starburst triggered during galaxy mergers. 
The results of the models with random gas directions are discussed in Appendix~\ref{section:incoherence}.

\section{Results}
\label{section:result}

\subsection{General properties of the gas flow near BHs from the simulations}

Before discussing in detail the results of the spin models, we briefly describe the properties of the gas flows near BHs and how they come to influence the angular momentum of the BHs. These results are derived directly from the simulations, before any post-processing is applied.  In Fig.~\ref{fig:BHexample}, ~\ref{fig:BHexample_z0-0.2} and ~\ref{fig:BHexample2} we show some individual BH histories that highlight the main findings. In this section we focus only on the two bottom panels, that show the degree of temporal coherence of the gas angular momentum through $\rm j_{\rm gas, z}$, which is the cartesian component along the (fixed) z-axis of the simulation box, of the gas angular momentum unit vector, and the accretion rate in Eddington units, $\chi$. 

Fig.~\ref{fig:BHexample} shows the evolution of a BH hosted in what becomes by $z=0$ a central galaxy of a cluster of galaxies.  The BH mass is $5\times10^9\, \rm M_\odot$ and lies in a host galaxy with a stellar mass of $9.7\times 10^{11}\, \rm M_\odot$, and host halo mass of $M_{\rm vir}=1.6\times 10^{14}\, \rm M_\odot$. At high redshift, $z>2$, the BH regularly accretes gas at or close to its Eddington limit. The gas possesses a high degree of coherency, with small variations of the degree of alignment with the galaxy total angular momentum. At low redshift, $z<2$,  the galaxy has consumed most of its gas through star formation, and blown some gas away through feedback (mostly due to AGN feedback in such a massive elliptical). Fig~\ref{fig:BHexample_z0-0.2} highlights this new phase, by zooming in on $0<z<0.2$ for the same BH. The only available gas is  hot plasma, at temperatures of a few $10^7\, \rm K$ and densities of $0.01$-$0.1\, \rm cm^{-3}$, that does not settle  into a well structured disc. The rapid variations of the gas angular momentum direction are driven by the local turbulence and bear no connection to the rotation axis (if any) of the galaxy. Accretion rates at low redshift can vary by several decades from a few $\chi=10^{-6}$ over long periods of time up to $\chi=1$ with short burst durations. In general, however, the mean accretion rates are very low (except for brief bursts due small, short-lived cooling flows). The tiny accretion rates are dictated by the absence of any cold star-forming gas in the galaxy. Gas accretion, in this phase, does not contribute significantly to the BH evolution, neither it mass, nor its spin, as detailed below. 

In Fig.~\ref{fig:BHexample2}, we show instead the evolution of a massive  disc galaxy of $M_{\rm s}=5\times 10^{10}\, \rm M_\odot$ in a $M_{\rm vir}=2\times 10^{12}\,Ê\rm M_\odot$ halo hosting a $M_{\rm BH}=10^8\, \rm M_\odot$ BH  at $z=0$. This galaxy lies in the field (its host halo is not a sub-halo of a more massive halo) and  contains in permanence a reservoir of cold star-forming gas supplying accretion for its central BH (and for star formation). In this case we see that the gas maintains a coherent angular momentum down to low-redshift, while at the same time the reservoir of cold gas keeps on feeding the BH at relatively high-levels. The bottom panel shows a general negative gradient of the accretion rate with decreasing redshift (from $\chi \sim 0.1$ at $z=3-4$ to $\chi \sim 10^{-3}$ at $z<1$), highlighting that  gas is still being consumed over cosmic time. Overall, it is gas accretion that drives this BH's evolution, both its mass, and its spin. 

These properties of the galaxy gas can help us understand how its angular momentum, temperature and density affect (or in some case do not affect) BH mass and spins. To assess the robustness of these results we describe in Appendix~\ref{section:zoom-halo}  two simulations at much higher resolution (10 pc), one of a high-redshift halo in a cosmological simulation, the other of an isolated halo. The former allows us to estimate the effects of galaxy mergers and of cosmic feeding, the latter allows us to highlight the specific influence of AGN feedback and of the dynamics of gas clumps, all this down within nuclear scales. In all cases we find that the gas angular momentum at scales of tens of parsec shows a significant degree of correlation with the gas angular momentum on  galactic scales, thus validating our assumptions. 

\subsection{Examples of BH mass and spin evolution}

In order to understand the connection between the BH mass growth and spin, we inspect the redshift evolution of the two BHs described in section 4.1. Fig.~\ref{fig:BHexample}  shows one of the most massive BHs in our simulation, hosted at $z=0$ in the central galaxy of a cluster of galaxies. 
Fig.~\ref{fig:BHexample} shows the BH mass as a function of redshift with its related spin evolution, angle between BH spin and gas angular momentum, the z-cartesian component of the BH and gas normalised angular momenta, and the BH accretion rate in Eddington units.
At high redshift, $z>2$, the BH steadily accretes gas close to its Eddington limit,  with strong coherence of the  accreted gas angular momentum, because the direction of the gas angular momentum with redshift does not vary much. In this situation, the large values of $\chi$ force the BH to re-align with its surroundings ($\cos \theta \sim 1$) and the BH spins at its maximum $a\sim 1$. Mergers between BHs (and their host galaxies) are present in the high redshift evolution, and, if BH mass ratios are sufficiently large, they change the value of the BH spin (e.g. at $z=2.9$ from $a\sim1$ before merger to $a\sim 0.8$ after merger).
However, as mergers are still extremely gas rich, a rapid accretion of cold gas perdures after the coalescence of the two BHs, and the spin value gets soon back to its maximum $a=0.998$. 

The mechanism at low redshift ($z<2$) is orthogonal to the picture describing the spin evolution at high redshift. As gas is depleted through star formation and feedback, spin decouples from the surrounding gas angular momentum.  The angle between the two angular momentum directions (gas and BH) strongly varies with redshift, essentially due to rapid variations of the gas angular momentum direction driven by the local turbulence of the hot plasma (see Fig.~\ref{fig:BHexample_z0-0.2}). Conversely, the BH spin direction is much more stable with long periods of time without changing direction (several hundreds of Myr) because the rates of gas accretion onto the BHs are too small to realign the BH spin with the local gas angular momentum.  Since for high-mass BHs the total angular momentum of the BH+disc system, $\bmath{J}_{\rm tot}$, is dominated by the BH's spin, therefore, although the direction of $\bmath{J}_{\rm d}$ changes erratically, the re-orientation of the BH is limited, as its spin direction is already close to the direction of $\bmath{J}_{\rm tot}$ \citep[cf.][]{dottietal13}.  The rapid changes in spin magnitude and directions are due either to BH coalescences that redistribute the final BH spin direction (e.g. at $z=0.2$ where a merger with BH mass ratio of 1:2 happens), or to the formation of a small cooling flow of gas onto the central galaxy (e.g. at $z=1$ where $\chi\gg10^{-5}$ but no mergers are detected).

In Fig.~\ref{fig:BHexample2} we show the results for a $M_{\rm BH}=10^8\, \rm M_\odot$ BH within a massive  disc galaxy. This field  galaxy experiences  very few mergers (the fraction of BH mass gained by coalescences is 6 per cent) and remains rich in cold gas down to $z=0$.  The difference with the previous BH is striking, the BH maintains a spin parameter close to its maximum $a=0.998$ because of the remarkable absence of significant mergers (except at $z=3.4$) and the moderate levels of accretion ($\chi\simeq 10^{-3}$-$10^{-2}$) maintained over a Hubble time in a coherent fashion (direction of the gas angular momentum varies little with time).

\begin{figure}
\vspace{-0.5cm}
  \centering{\resizebox*{!}{6.2cm}{\includegraphics{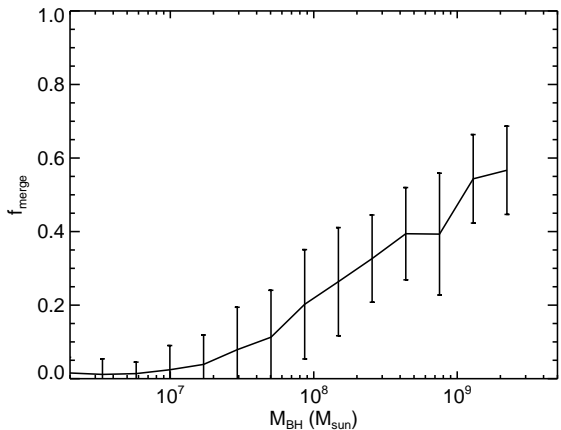}}}
  \caption{Average fraction of BH mass gained through coalescence with companions as a function of the BH mass at z $=0$. Error bars correspond to the standard dispersion. Contribution from BH coalescence to the final mass is more important for the most massive BHs. Lower mass BHs gain mass preferentially through gas accretion from their environment. }
    \label{fig:fmergebh_ave}
\end{figure}

\begin{figure}
\vspace{-0.5cm}
  \centering{\resizebox*{!}{6.2cm}{\includegraphics{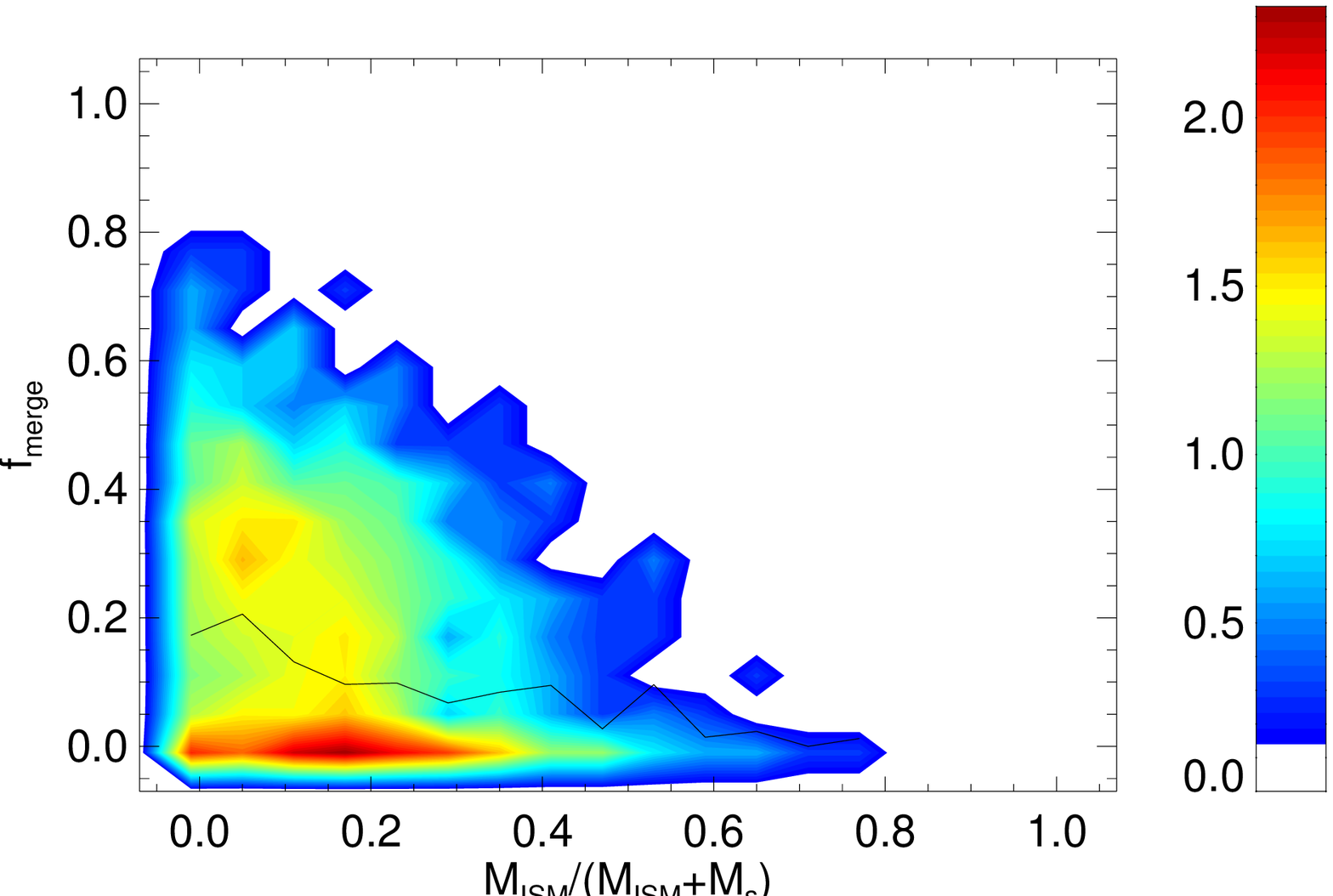}}}
  \centering{\resizebox*{!}{6.2cm}{\includegraphics{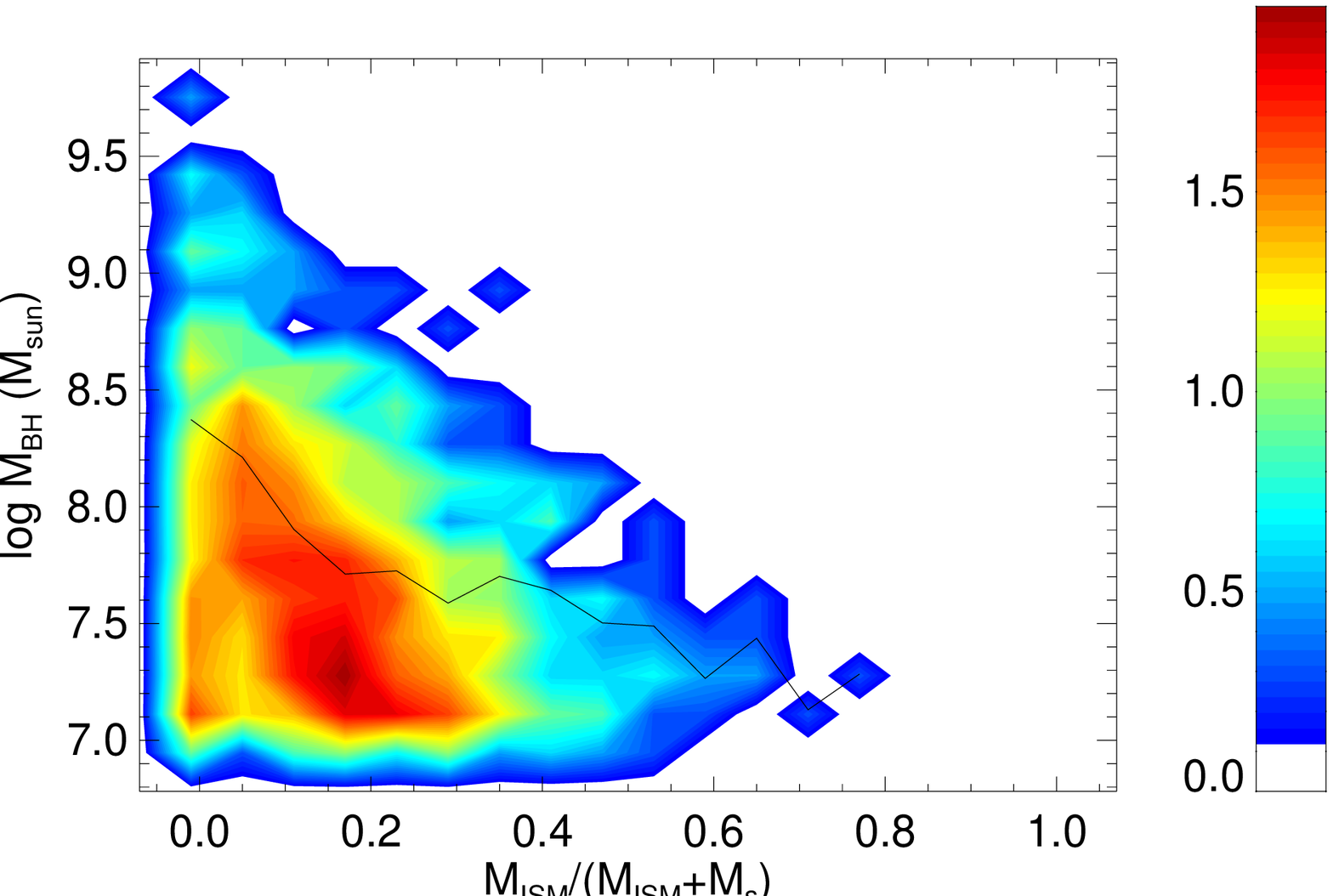}}}
  \caption{Number-weighted distribution of BH mass fractions gained through coalescence (top) and of BH mass (bottom) as a function of the gas fraction in their host galaxy at $z=0$ for BHs with masses above $M_{\rm BH}>10^7\, \rm M_\odot$. Solid lines are the average of the distribution. BHs whose mass is mostly gained through coalescence instead of direct gas accretion are preferentially hosted in gas-poor galaxies.  BHs which have not gained much mass though coalescences hosted in gas-poor galaxies are the BHs which grow the least.  }
    \label{fig:fmergebh_cont}
\end{figure}

%
\subsection{BH mass growth and spins and their relation to gas accretion and mergers}
\begin{figure}
  \centering{\resizebox*{!}{6.2cm}{\includegraphics{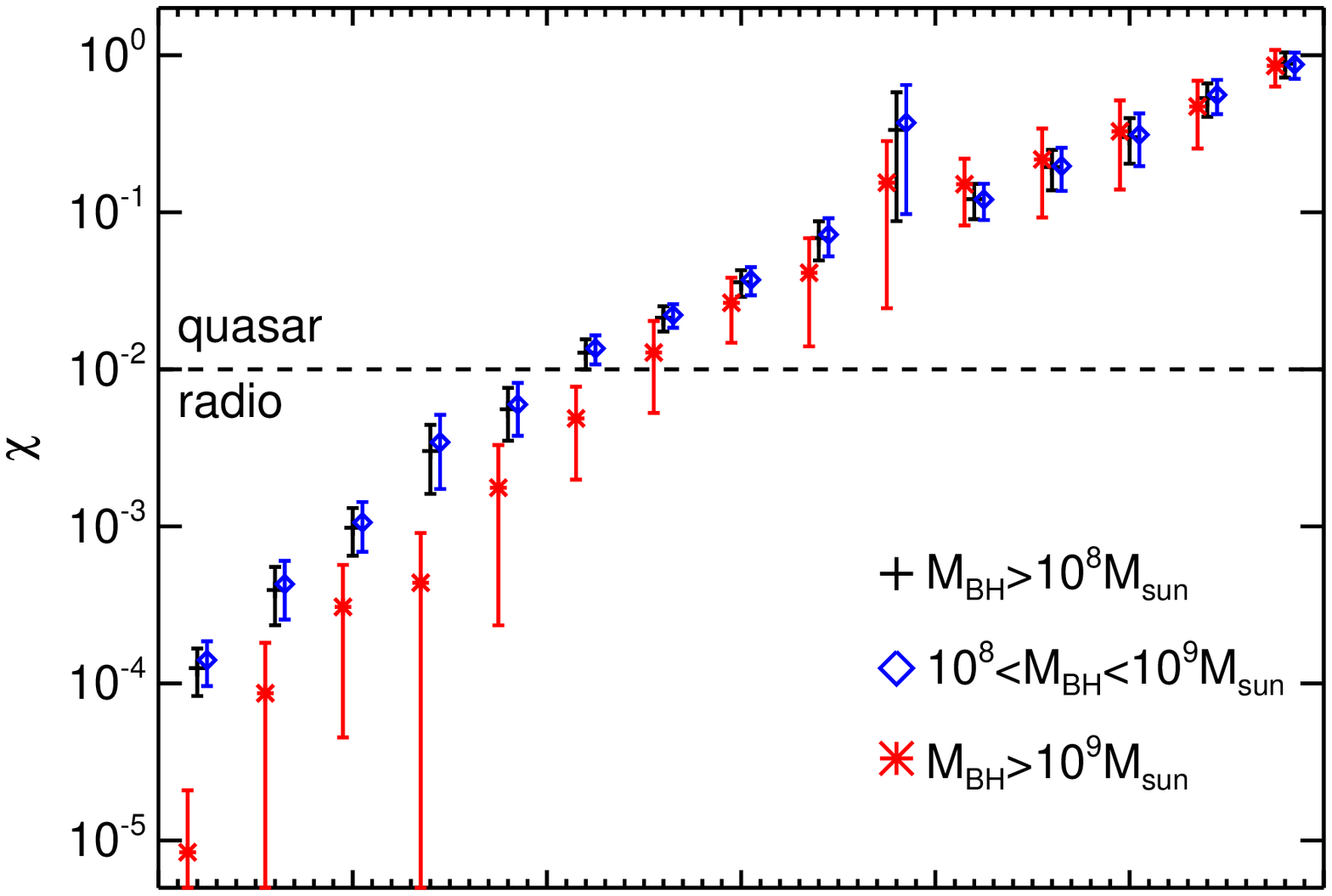}}}\vspace{-1.3cm}
  \centering{\resizebox*{!}{6.2cm}{\includegraphics{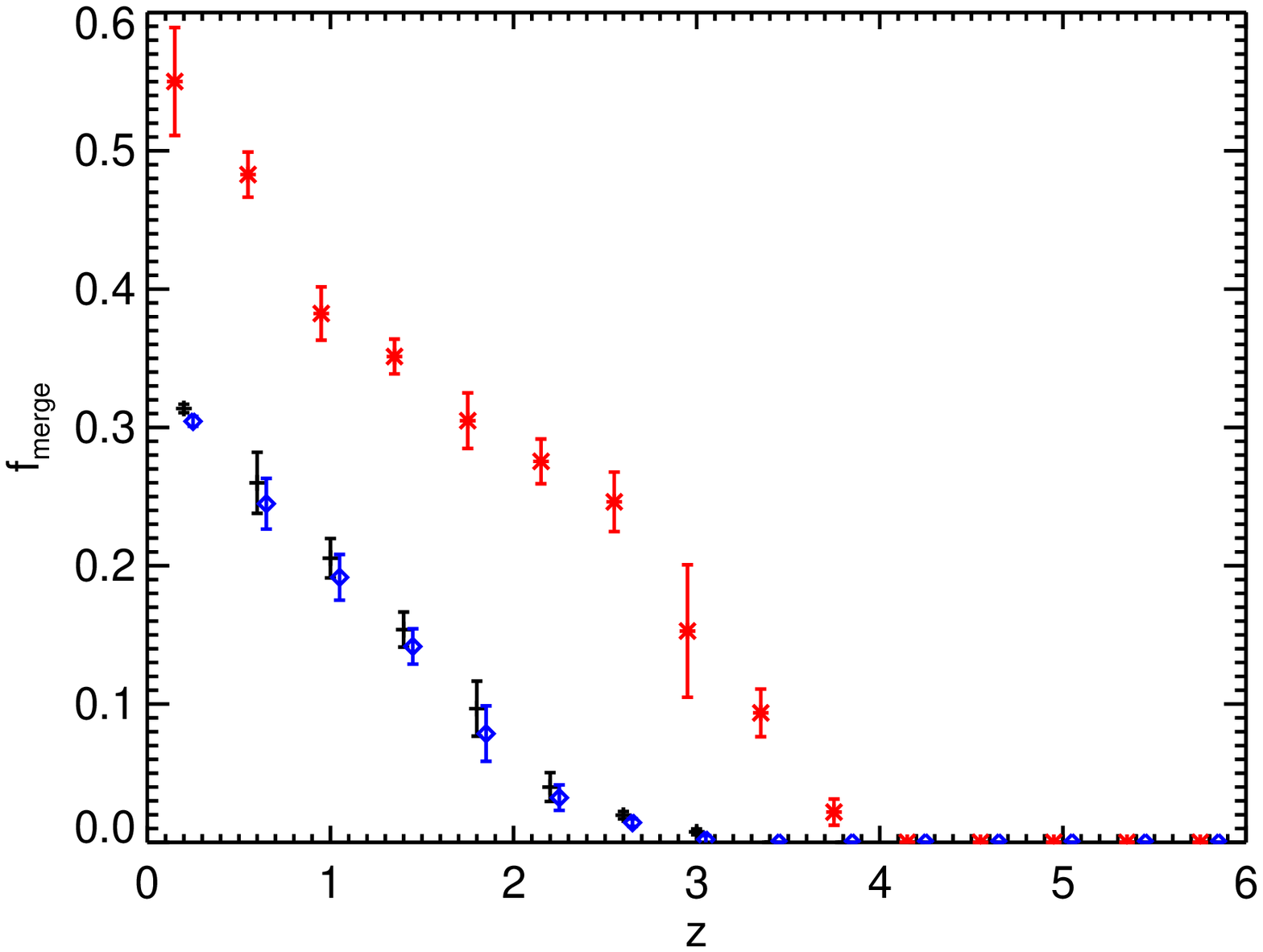}}}
  \caption{\emph{Top}: Average value of the Eddington ratio as a function of redshift for selected BHs with masses at $z=0$ above $M_{\rm BH}\ge10^8\, \rm M_\odot$ (black), $M_{\rm BH}\ge10^9\, \rm M_\odot$ (red), and $10^8 \le M_{\rm BH}\le10^9\, \rm M_\odot$ (blue) with their standard dispersion (error bars). 
\emph{Bottom}: Average value of the fraction of BH mass gained through mergers.
  The average accretion rate onto BHs is steadily decreasing over time, and at redshift $z\sim 2$ the most massive BHs switches from a quasar mode ($\chi \ge 0.01$) to a radio mode ($\chi < 0.01$). The contribution of mergers to the BH mass is increasing with time (decreasing redshift), starts to be significant by $z=2$-$3$, and the most massive BHs have the largest mass fraction of mergers. }
    \label{fig:chivsz}
\end{figure}

The two  BHs described in section~4.2 are extreme cases with one essentially built-up by mergers and the other one built-up by accretion. Indeed, it exists a  range of BHs with moderate fractions of mass gained through mergers for different values of BH masses. In the following, we estimate the relative contribution of gas accretion and BH coalescences to the build-up of the BH population, and pinpoint how the mass built-up of BHs is expected to determine the spin of these compact objects.

\subsubsection{Accretion and mergers: mass growth}

We measure the fraction of BH mass that has been gained through coalescences with companions (as opposed to growth by direct gas accretion) $f_{\rm merge}=\Delta M_{\rm merge}(<t)/M_{\rm BH}$, where $\Delta M_{\rm merge}(<t)$ is the total integrated mass gained by mergers for a given BH, as a function of BH mass at $z=0$ in Fig.~\ref{fig:fmergebh_ave}.
Note that this definition does not take into account the fact that the merged companion of a given BH has grown by direct gas accretion, all its mass before coalescence is added to the merged mass $M_{\rm merge}$.
Low mass BHs $M_{\rm BH}\lesssim10^8\, \rm M_\odot$ have gained most of their mass through direct accretion of gas, $f_{\rm merge}<0.2$, while the most massive BHs are dominated by mergers, $f_{\rm merge}>0.5$ for BHs with $M_{\rm BH}>10^9\, \rm M_\odot$ \citep[see also][]{Malbon2007,Volonteri2013,kulieretal13}.

We can also estimate the correlation between the amount of gas in galaxies and the channel through which their BHs have gained most of their mass. We define the ratio of gas in the galaxy by $M_{\rm ISM}/(M_{\rm ISM}+M_{\rm s})$, where $M_{\rm s}$ is the total stellar mass of the galaxy obtained with a halo finder (AdaptaHOP, see~\citealp{aubertetal04, tweedetal09}) performed on star particles, and $M_{\rm ISM}$ is the star forming gas within the galactic radius with gas densities $n > 0.1 \, \rm cm^{-3}$. We include ony BHs with masses above $M_{\rm BH}>10^7\, \rm M_\odot$ to avoid resolution effects.  Most of the BHs that gain 50 per cent or more of their mass through coalescences inhabit galaxies with low gas fractions $M_{\rm ISM}/(M_{\rm ISM}+M_{\rm s})<0.2$ (top panel of Fig.~\ref{fig:fmergebh_cont}), but we see that BHs in galaxies with very low gas-content have not necessarily all grown in mass by mergers (see the red plume with $f_{\rm merge}<0.1$ and $M_{\rm ISM}/(M_{\rm ISM}+M_{\rm s})<0.2$ on the lower-left corner).  These BHs, which have not gained much mass though coalescences and they are hosted in gas-poor galaxies, are the BHs which grow the least (i.e. are the least massive ones in the distribution, see bottom panel of Fig.~\ref{fig:fmergebh_cont}). 

Fig.~\ref{fig:chivsz} shows the evolution of the Eddington ratio with redshift for a sub-sample of BHs\footnote{A glitch in the Eddington ratio is visible at $z=4$ and is due to one extra level of refinement triggered at that moment, increasing the force sampling in galaxies and, thus, the density of the cold gas, which in turns produce larger values of the gas accretion rate.}.
We selected BHs with masses above $10^8\, \rm M_\odot$ at $z=0$ and followed their progenitors at higher redshift.
The progenitors of these massive BHs accrete gas close to the maximum Eddington rate at high redshifts, and release their energy into quasar mode.
At $z\sim 2$ these BHs switch from a quasar mode to a radio mode of AGN feedback ($\chi < 0.01$), and the Eddington ratio reaches $\chi \sim 10^{-5}-10^{-4}$ at $z=0$, which correspond to a typical growth rate of $t_{\rm Edd}/\chi\sim 1000$ Gyr, where $t_{\rm Edd}=45$ Myr is the e-folding time of a BH accreting at the Eddington rate (assuming $\epsilon_{r}=0.1$).
It means that, at low redshift, the most massive BHs stop growing by accretion of gas from their surroundings, they can eventually get more massive by mergers.
Indeed, the fraction of mass gained through mergers is continuously increasing with time, and represents 30 per cent of the mass built-up of BHs at $z=0$ for those with $M_{\rm BH}>10^8\, \rm M_\odot$, as we discussed above (cf. Fig~\ref{fig:fmergebh_ave}).

An interesting point is that more massive is the BH the faster the quenching of gas accretion (relative to Eddington) is reached, e.g. progenitors of $10^9\, \rm M_\odot$ BHs at $z=0$ have lower Eddington ratio than progenitors of $10^8\, \rm M_\odot$ BHs.
The same is true for the fraction of mass gained through mergers: mergers start to be significant ($f_{\rm merge}>0.1$) earlier (at $z\simeq3$ for $M_{\rm BH}>10^9\, \rm M_\odot$ instead of $z\simeq2$ for $10^8\, \rm M_\odot<M_{\rm BH}<10^9\, \rm M_\odot$), and represent a more significant fraction of the mass growth at $z=0$ ($f_{\rm merge}\simeq 0.55$ instead of $0.3$).

\begin{figure}
  \centering{\resizebox*{!}{3.3cm}{\includegraphics{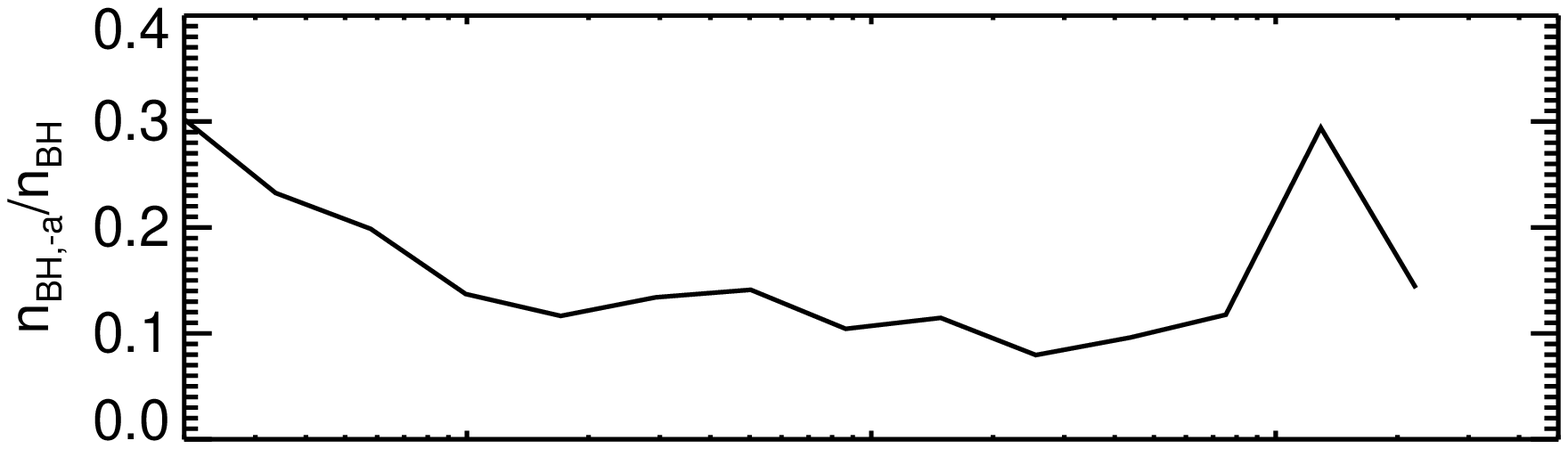}}}\vspace{-1.3cm}
  \centering{\resizebox*{!}{6.2cm}{\includegraphics{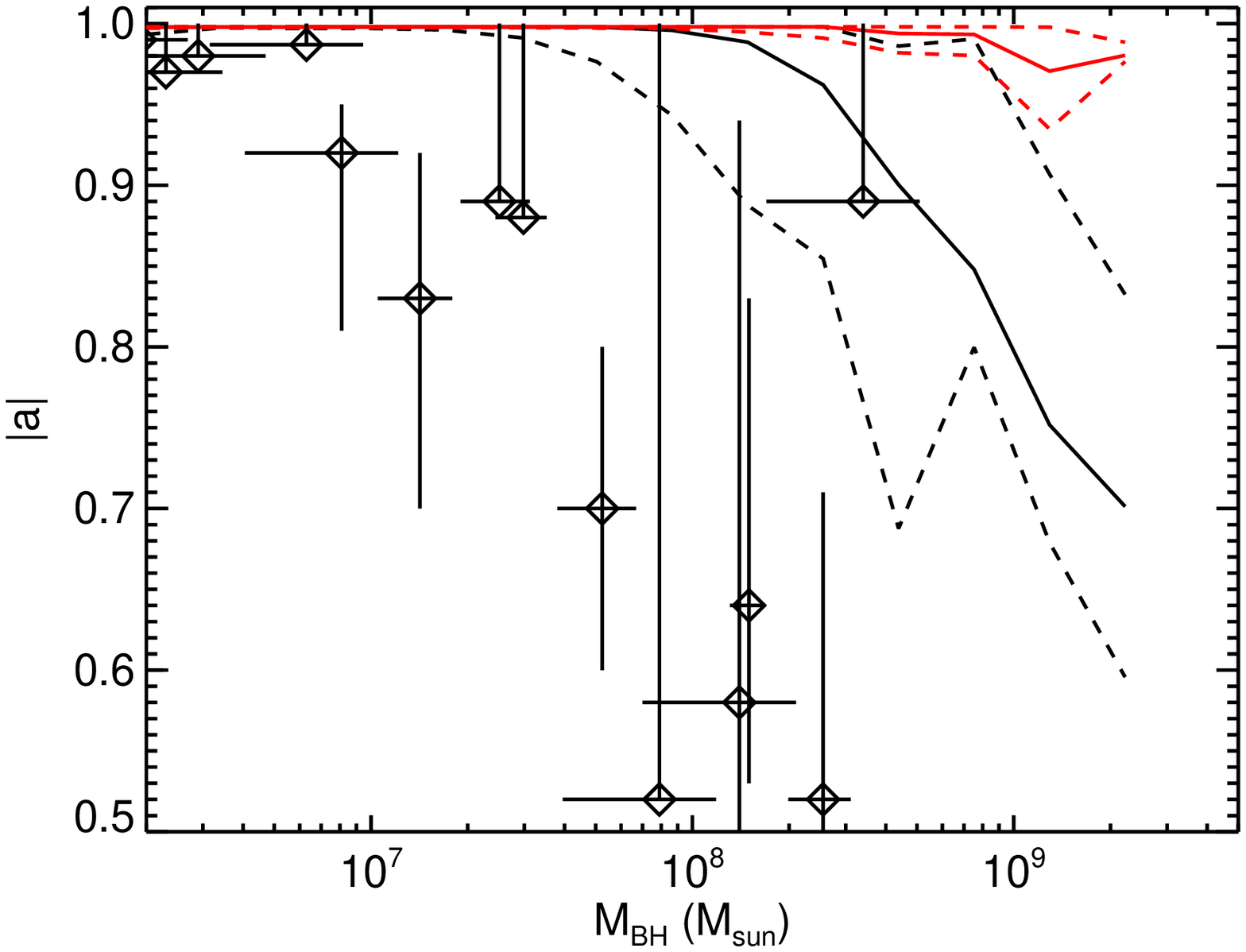}}}
  \caption{\emph{Top}: Average fraction of BHs with negative spins as a function of the BH mass; \emph{bottom}: Median of the absolute value of the spin as a function of the BH mass (solid line), with 20 and 80 percentile values (dashed lines). The diamonds with $1\sigma$ error bars are a compilation of data points with a quality assessment from~\citet{reynoldsetal13}. Measurements in the simulations are made at $z=0$. The red curve corresponds to the median of the absolute value of the spin for a model of spin evolution where BH coalescence does not modify the spin of the remnant.}
    \label{fig:abhvsmbh_ave}
\end{figure}

\begin{figure}
  \centering{\resizebox*{!}{6.2cm}{\includegraphics{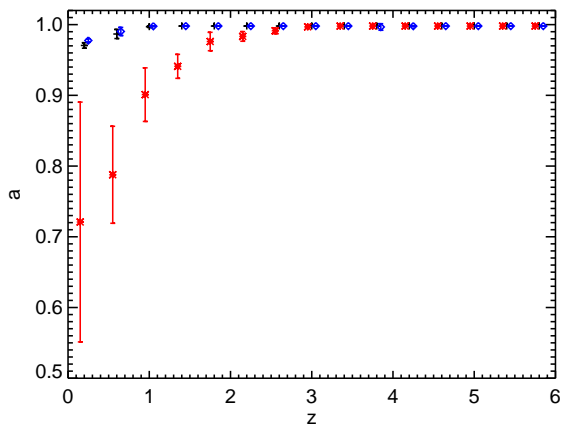}}}
  \caption{Same as Fig.~\ref{fig:chivsz} for the average spin parameter. Spins are maximum at high redshift when accretion drives the growth of BHs and the spins decrease at low redshift when binary coalescence start to contribute significantly to the final BH mass. }
    \label{fig:abhvsz}
\end{figure}

\begin{figure}
  \centering{\resizebox*{!}{6.2cm}{\includegraphics{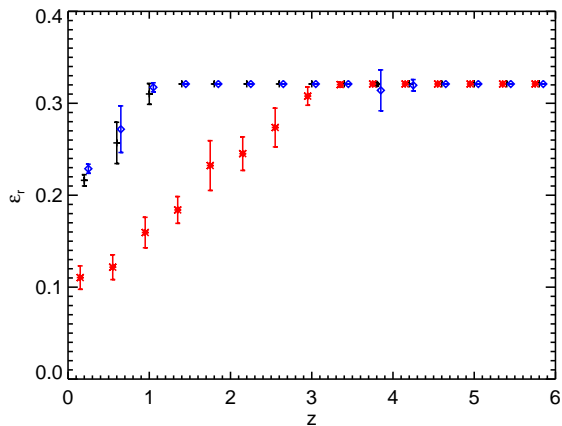}}}
  \caption{Same as Fig.~\ref{fig:chivsz} for the average radiative efficiency. Radiative efficiencies follow the evolution of the spins with redshift, with maximum efficiencies at high redshift and reduced efficiencies at low redshift because of lower spin values.}
    \label{fig:eradvsz}
\end{figure}

\subsubsection{Accretion and mergers: spin evolution}

Fig.~\ref{fig:abhvsmbh_ave} shows the spin value as a function of the BH mass at $z=0$. Note that BHs with masses below $10^6\, \rm M_\odot$ are not represented, because we consider that they are too close to the chosen initial seed BH mass, and they are still affected by the mass resolution sampling.
The least massive BHs we consider ($ 10^6 \, \rm M_\odot \lesssim M_{\rm BH} \lesssim 10^7 \, \rm M_\odot$) have large spins, close to $|a|=0.998$, while the most massive BHs ($M_{\rm BH}\geqslant 10^8 \, \rm M_\odot$) tend to lower spin values, with $| a |\simeq0.7-0.8$ for BHs with mass $M_{\rm BH}\simeq 10^9 \, \rm M_\odot$.  Coherent gas accretion onto BHs, i.e. gas angular momentum aligned with BH spin, is the reason for the spin values approaching $a=0.998$: gas in disc galaxies is set-up on almost steady orbits, and contributes constructively to the spin amplification.
Note that a non-negligible fraction of BHs, between 10 and 20 per cent, have a negative spin value that also tends to $a=-0.998$ for low mass BHs and to a more intermediate value $a \simeq -0.7$ for the most massive BHs.
Midsize mass BHs ($M_{\rm BH}\lesssim 10^8 \, \rm M_\odot$) with negative spins have the gas angular momentum vector flipped with respect to the  spin direction  because feedback from the central AGN removes gas from the central region, while at the same time their typical levels of accretion are sufficiently low that the magnitude of the spins does not change to intermediate spin negative values through accretion of counter-rotating material. 
At larger masses, spin magnitudes deviate significantly from unity and this may be due to either: (i) gas angular momentum decoherence with respect to the BH spin or (ii) coalescence of BHs.

Coalescences of BHs can rapidly change the BH spin magnitude and direction.  Comparing Figs.~\ref{fig:fmergebh_ave} and~\ref{fig:abhvsmbh_ave} it appears that the most massive BHs at $z=0$ that have grown mostly by coalescences have lower spin values, and thus, galaxy mergers at low redshift that carry little gas, are responsible for spinning down BHs. We tested this scenario by deactivating the spin evolution during BH coalescence, i.e. BHs mass is still updated but not their spin, and we see in Fig.~\ref{fig:abhvsmbh_ave}, that the red curve is extremely close to $a=0.998$ even for the most massive BHs.
It clearly demonstrates that BH mergers are responsible for spinning down BHs at low redshift and flipping the spin direction.

Despite the intrinsic difficulty in directly measuring BH spins from observations, several attempts have been made  through X-ray spectroscopy.  Contradictory results~\citep{patricketal11, brennemanetal11, reynoldsetal12} exist, mostly because of different approaches in modelling the continuum of the spectral energy distribution. One recent measurement, obtained using NuSTAR high-energy capabilities, suggests that the BH with mass $2\times 10^6 \, \rm M_\odot$ in the centre of NGC 1365 is spinning at its maximum value~\citep{risalitietal13}.

We included in Fig.~\ref{fig:abhvsmbh_ave} the results from observations for data points that passed a given quality assessmentõ\citep{reynoldsetal13}. While we see the same trend in data and pseudo-data, i.e. decreasing spin with increasing BH mass, observed spins tend to be slightly smaller than simulated ones. There is a possibility that due to finite mass resolution effects, the small mass satellites are not resolved and thus the amount of accretion over merger is over-predicted, favouring larger spin values. However, to change significantly the value of a spin during coalescence, the mass ratio has to be significant $q\gtrsim0.1$ (see Fig.~\ref{fig:BHexample} for an illustrative example), therefore, significant BH-BH coalescences for redistributing spins are already resolved. Another possibility, which is not resolution-dependent, is that there is still too much gas in galaxies during mergers, so that after decreasing the BH spin after a merger, the BH remnant spins up because of the too large reservoir of gas and revived accretion onto the BH. 

We compute the evolution of the average spin parameter as a function of redshift for the same sample of supermassive BHs (see Fig.~\ref{fig:abhvsz}).
At high redshift $z>2$, the progenitors of the most massive BHs at $z=0$ have spin close to their maximum value $a=0.998$.
At lower redshift (below $z<1-2$), the value of the spin decreases with time as BHs experience more mergers that change spin directions and spin down BHs.
As gas accretion onto BHs is quenched efficiently at low redshift, gas cannot spin back up the BHs efficiently.

From the values of spin, it is possible to evaluate the radiative efficiency (which is kept as a constant in our feedback model and equal to $\epsilon_{\rm r}=0.1$) of the accretion process which writes $\epsilon_{\rm r}=1- E_{\rm isco}=1-\sqrt{1-2/(3r_{\rm isco})}$, where $E_{\rm isco}$ is the specific energy of a gas particle in the innermost stable circular orbit.
Thus, for a maximum spin of $a=0.998$, the radiative efficiency is equal to $\epsilon_{\rm r}=0.321$ ($\epsilon_{\rm r}=0.057$ for $a=0$ and $\epsilon_{\rm r}=0.038$ for $a=-0.998$), which is the exact value reached for BHs at high redshift when they are maximally spinning (see Fig.~\ref{fig:eradvsz}).
While spins are decreasing with redshift, the radiative efficiencies are also decreasing, reaching $\epsilon_{\rm r}=0.11$ for $M_{\rm BH}> 10^9 \, \rm M_\odot$ at $z=0$ and larger values for less massive BHs.

The high values of spin, and, thus, of radiative efficiencies measured at high redshift have possible important consequences for the growth of high-redshfit BHs, as the most massive ones need to grow by several orders of magnitude to reach $10^9\, \rm M_\odot$ at $z=6$ in less than a billion year to explain observed powerful quasars in the high redshift Universe.
Given the timescale, they need to accrete gas at the Eddington rate, but given an efficiency of $\epsilon_{\rm r}= 0.321$ the growth time scale is $t_{\rm Edd} (\epsilon_{\rm r}=0.321)=145\, \rm Myr$ (to be compared to 45 Myr for $\epsilon_{\rm r}=0.1$), and, thus would require a seed mass BH of $10^6\, \rm M_\odot$ possibly somewhat unrealistic.  If the spin is so large even at high redshift, this would favour a super-Eddington mode of accretion, or that  the  efficiency in producing radiation is decreased by, e.g.,  jet production \citep{Ghisellini2013}.

Some empirical constraints from~\cite{Li2012} and~\cite{Shankar2013}, using continuity equation approaches, suggest that the average radiative efficiency decreases with decreasing redshift (in agreement with these results).  However,~\cite{Shankar2013} favours lower typical radiative efficiencies, in general, between 0.05 and 0.14~(see also \citealp{merloni&heinz08}). This models also prefer an increasing radiative efficiency with increasing BH mass (opposite to our results). 
On the other hand,~\cite{Li2012} find almost no dependence of the radiative efficiency with BH mass at $z<1$ and an increasing radiative efficiency with mass at $z>1$ (see their Fig.~6). 
Regarding the redshift dependence, their results for the most massive BHs ($\log M_{\rm BH}>8.5$) is in perfect agreement with ours, while at lower masses ($7.5<\log M_{\rm BH}<8.5$) the result is opposite to ours: no redshift dependence, and low values of the radiative efficiency, $\epsilon_r \sim0.05$. 

Finally, \cite{2011MNRAS.414.1937M}  use jet efficiencies from magnetohydrodynamic simulations of jets coupled with the X-ray luminosity function and the local mass function of BHs  to infer the spin distribution and its evolution. Their model favours a bimodal spin distribution, very little cosmic evolution, and that the most massive BHs have decreased their spin over cosmic time \citep{2011MNRAS.418L..84M}. Conversely, \cite{2011MNRAS.414.1253D} link directly the radio power of sources to the BH spin and find that there is no strong correlation between spins and BH mass (the sample has  $\log M_{\rm BH}>7.5$), and that spins tend to increase at higher redshift \citep[see also][]{2009ApJ...696L..32D,2013arXiv1312.4862D}.

\subsection{Spin magnitude and orientation with their host galaxy}

\begin{figure}
\vspace{-0.5cm}
  \centering{\resizebox*{!}{6.2cm}{\includegraphics{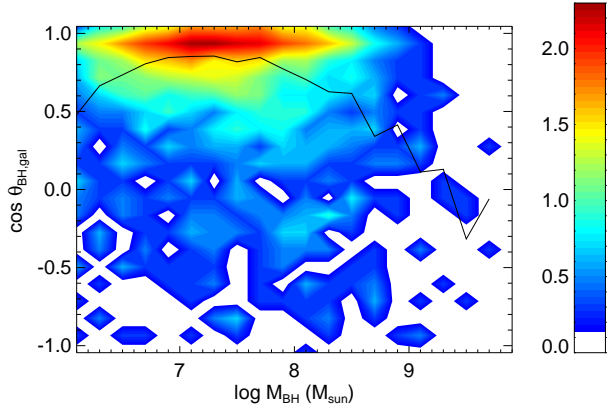}}}
  \caption{Number-weighted distributions of angles between BH spins and galactic angular momentum (defined by their stellar content) as a function of the BH mass at $z=0$. The solid line is the average of the distribution. Midsize mass BHs have spin aligned with their host galaxy, while the most massive BHs are randomly oriented. }
    \label{fig:anglebhgalvsmbh}
\end{figure}

\begin{figure}
\vspace{-0.5cm}
  \centering{\resizebox*{!}{6.2cm}{\includegraphics{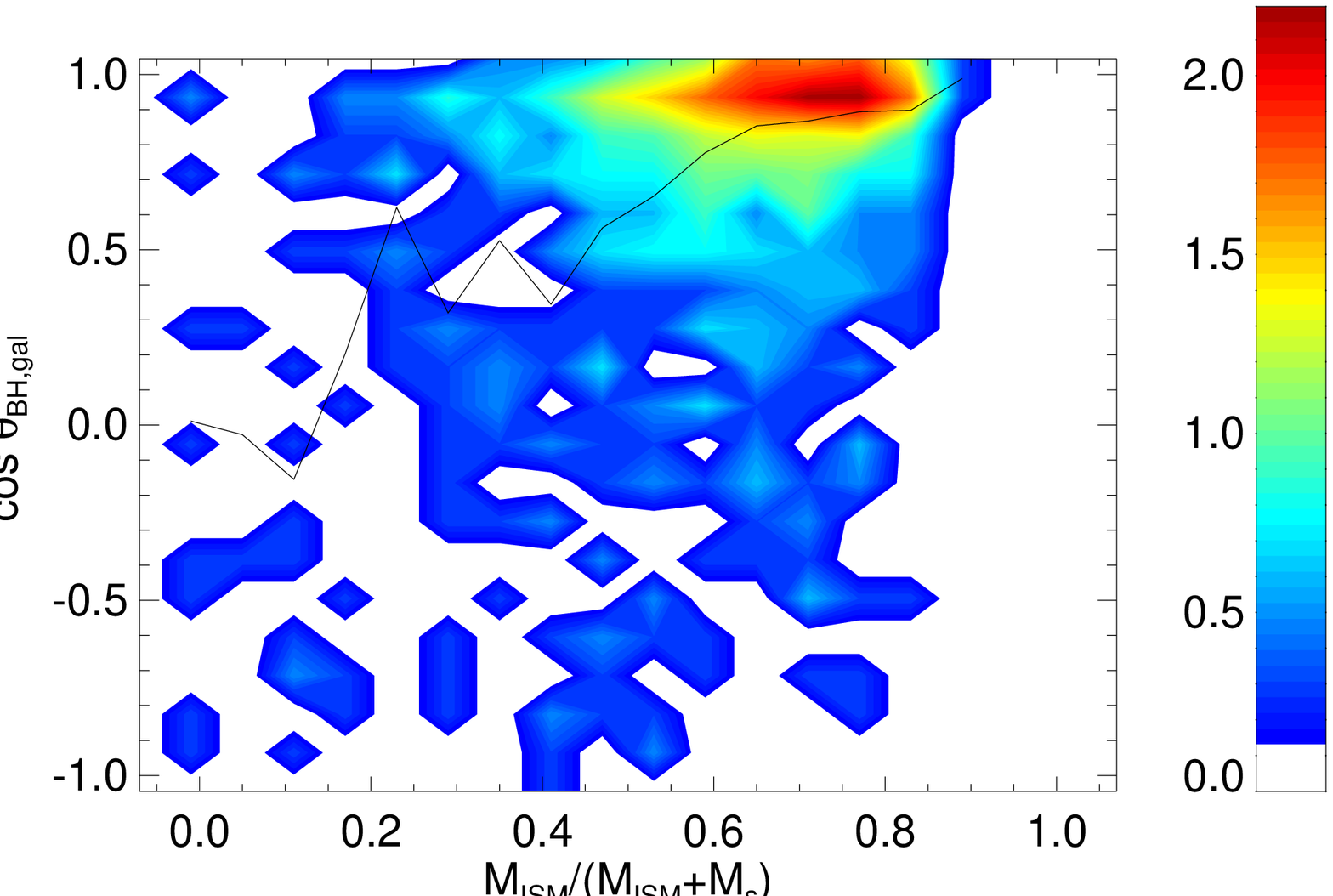}}}\vspace{-1cm}
  \centering{\resizebox*{!}{6.2cm}{\includegraphics{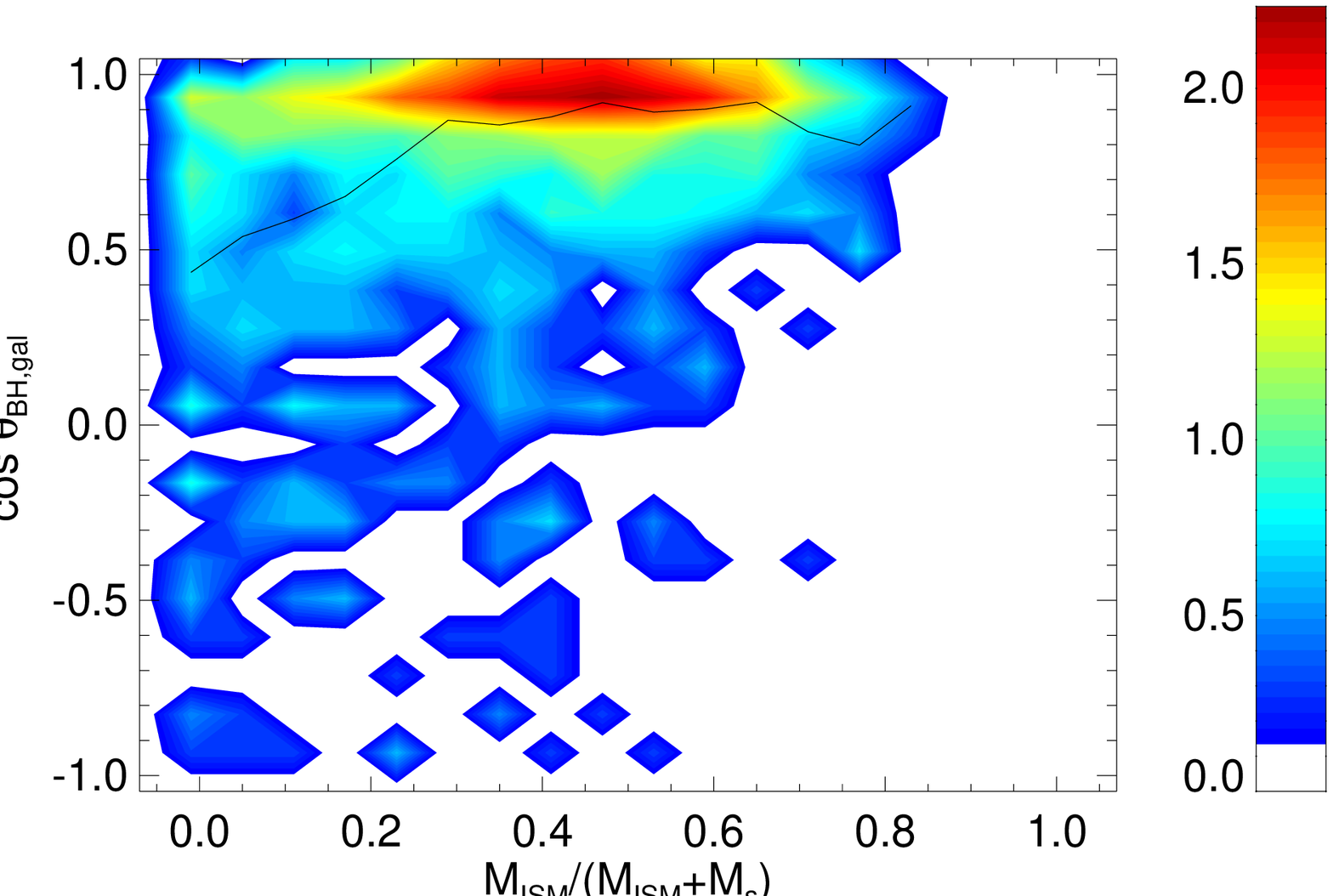}}}\vspace{-1cm}
  \centering{\resizebox*{!}{6.2cm}{\includegraphics{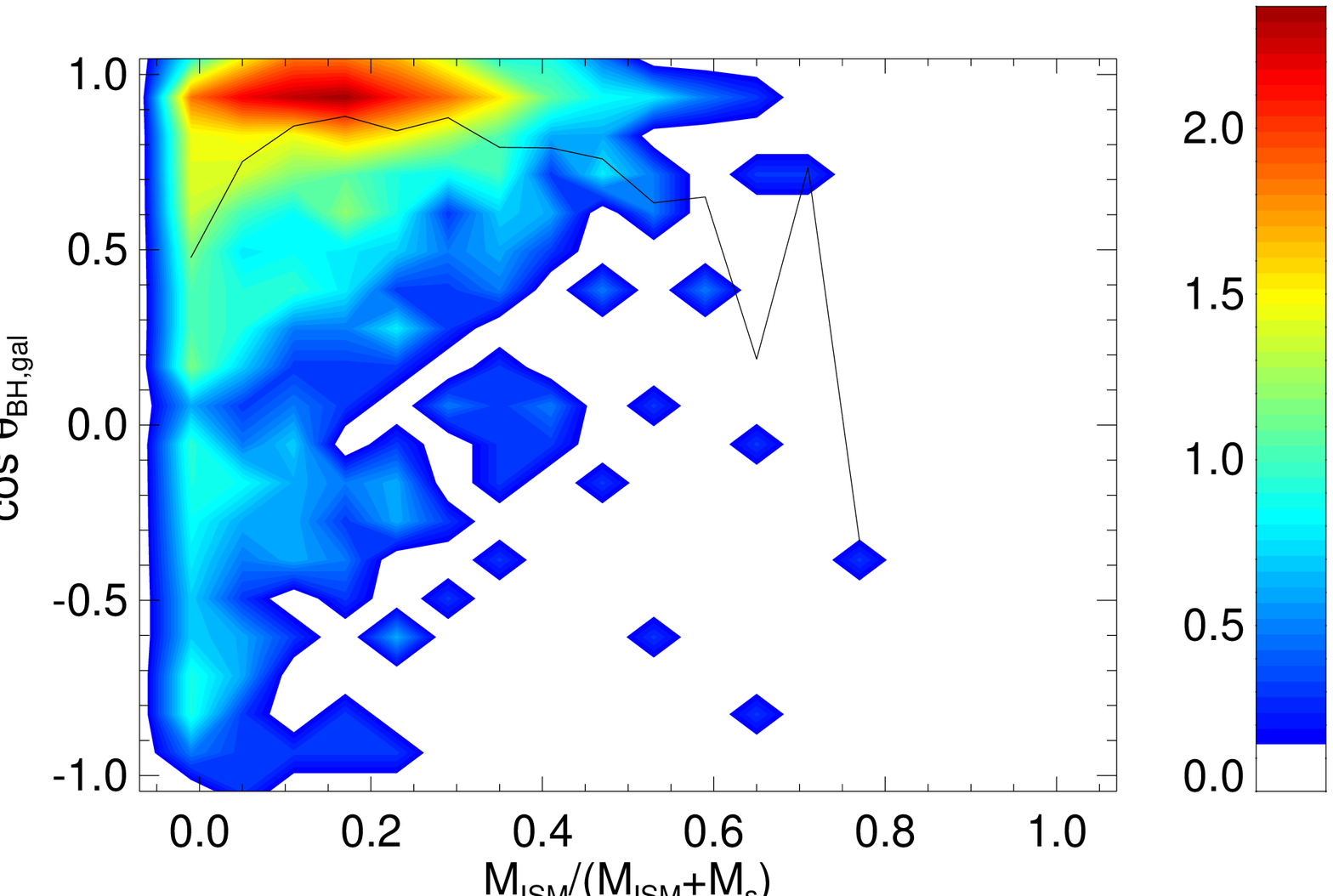}}}
  \caption{Number-weighted distributions of angles between BH spins and galactic angular momentum (defined by their stellar content) as a function of the gas fraction in their host galaxy at $z=2$ (top), $z=1$ (middle), $z=0$ (bottom). Solid lines are the average of the distribution. Gas poor galaxies host BHs whose spin is misaligned with the angular momentum of the stars (but not totally random), while gas-rich galaxies host strongly aligned BHs.}
    \label{fig:anglebhgalvsgasfrac}
\end{figure}

\begin{figure}
\vspace{-0.5cm}
  \centering{\resizebox*{!}{6.2cm}{\includegraphics{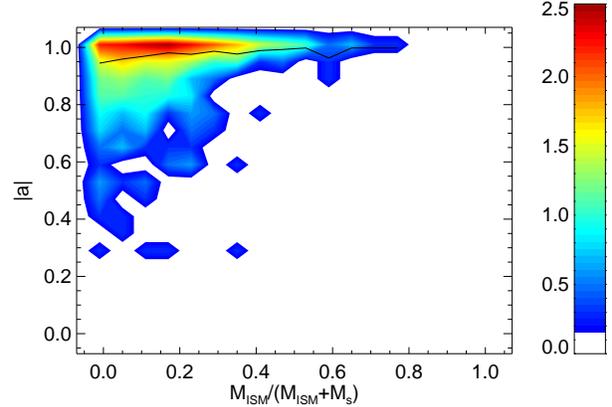}}}
  \caption{Number-weighted distribution of BH spins as a function of the gas fraction in their host galaxy at $z=0$. Solid lines are the average of the distribution. BHs whose mass is mostly gained through coalescence instead of direct gas accretion are preferentially hosted in gas-poor galaxies, which results in BHs with lower spin values.}
    \label{fig:spinvsgasfrac}
\end{figure}

We see that mergers are responsible for spinning down BHs, and that the most massive BHs with larger contribution from mergers to their mass are those with lower spins.
Fig.~\ref{fig:anglebhgalvsmbh} shows the distribution of angles between BH spins and the stellar angular momentum $\theta_{\rm BH,gal}$ of their host galaxy as a function of BH mass.
Midsize mass BHs with $10^7<M_{\rm BH}<10^8\,Ê\rm M_\odot$ show a high level of alignment with their host galaxy, as gas accretion is mostly responsible for both BH and stellar mass growth in this mass range. 
As BHs get more massive, spins are more randomly oriented with respect to their host galaxy angular momentum and tend towards $\cos \theta_{\rm BH,gal}=0$.
Mergers can rapidly change the orientation of galaxies, as do coalescences for BHs but they do not necessarily follow the same evolution while they merge.
BHs radiate a part of their orbital angular momentum in gravitational waves that  change the final orientation of the BH remnant relative the galaxy angular momentum.
Galaxies can also exchange angular momentum with their environment (gas and DM) by long-distance gravitational interactions.
Low mass BHs show a significant deviation from $\cos \theta_{\rm BH,gal}=1$, which is to be attributed to resolution effects.
A BH with mass  $M_{\rm BH}=10^6\, \rm M_\odot$ lie typically in host galaxy of $M_{\rm s}=10^9\,Ê\rm M_\odot$ which is composed of $\sim$ 50 stellar particles (while the hosts of BHs with mass $M_{\rm BH}=10^7\, \rm M_\odot$ are composed of $\sim$ 500 particles, and so forth).

We investigate if there is any correlation between the amount of gas in galaxies and the spin magnitude or orientation of their BHs.
Fig.~\ref{fig:anglebhgalvsgasfrac} shows the distribution of angles between BH spins and their stellar angular momentum $\theta_{\rm bh,gal}$ of their host galaxy as a function of the gas fraction in the galaxy for BHs with masses above $M_{\rm BH}>10^7\, \rm M_\odot$.
There is some correlation of $\theta_{\rm bh,gal}$ with the gas content in the galaxy: BHs in gas-rich galaxies have spins aligned with the galaxy angular momentum, and BHs in gas-poor galaxies have more randomly orientated spins though with some degree of correlation, $\cos \theta_{\rm bh,gal} = 0.5$ for the galaxies with zero star-forming gas.
Gas-rich galaxies provide both strong accretion onto BHs and actively star-forming regions through the same gas, thus contributing to aligning together BHs, stars and gas.
Vice versa, galaxies with low gas content are the systems less effective at realigning spins with stellar angular momentum, and these galaxies are those which host BHs whose fraction of mass gained through coalescence is the largest (see Fig.~\ref{fig:fmergebh_cont}).
However the orientation of spins is not purely random even for the gas-poor galaxies. 
In the case of the most massive galaxies, they  show some  degree of alignment because they experience intense mergers that efficiently reorientate spin and galactic angular momentum.  Lower mass galaxies, instead, have fewer mergers and keep the BH spin aligned with their stellar angular momentum as they both, the BH and the stellar component, grow from the same gas. 
Galaxies with low gas fractions exhibit lower degree of alignment with their central supermassive BHs at higher redshift.
This is essentially an effect of the quiescent depletion of gas: gas-rich galaxies at high redshift, that host strongly aligned BHs, consume their gas through the star formation process (and partially because of AGN feedback, see~\citealp{duboisetal12agnmodel}) and end up with lower gas fractions at low redshift, with BHs still aligned to their hosts if they have not endured a significant number of mergers.

We now turn to compare our models with the available observational constraints. \cite{BattyeBrowne2009} and \cite{BrowneBattye2010} find a higher degree of alignment between radio and optical emission for radio-quiet BHs hosted in bluer, less concentrated galaxies and a lower degree of alignment for redder, more concentrated galaxies and radio-loud objects.
Though, they do not find any specific dependence on stellar mass as we do (BH mass linearly correlates with stellar mass), their result is in broad agreement with our finding on spin orientation with gas richness.
 \cite{Orban2011} suggest that Narrow Line Seyfert 1s are fed through secular evolution and suggest how this should imply coherency in the angular momentum of the spin and of the disc.  On the other hand, the degree of alignment is very low to inexistent for Seyfert galaxies~\citep{Kinney2000, Middelberg2004, VerdoesKleijn2005, Gallimore2006}, which is in contradiction with our results.
However, the orientation of radio jets is not necessarily aligned with BH spins because the propagation of jets from the accretion disc up to galactic scale can be strongly perturbed by the multiphase structure of the interstellar medium.
As a consequence, gas-rich galaxies, which are more sensitive to gravitational instabilities with a clumpy cold gas structure, can deflect significantly the propagation of jets. Additionally, most well-studied sources with the Very Long Based Array (VLBA) are not disc galaxies as they are typically weak sources in radio, however, jet directions based on VLBA observations \citep{Kellermann2004} find kinks and bends on a variety of scales, as well as expected systematic effects such as precession, making it hard for available data to provide a reliable indication of the jet direction at the launching point, and its relationship with the axes of rotation of their hosts.  

Finally, we measured the distribution of BH spins as a function of the gas fraction of their host galaxy (Fig.~\ref{fig:spinvsgasfrac}).
Spins weakly correlate with gas fractions, gas-rich galaxies host BHs with maximum spin and gas-poor galaxies host BHs with spin values around $a=0.95$ slightly below the maximum.
The reason is that gas-poor galaxies can be galaxies where the merger fraction in the BH mass build-up is any between 0 and 1.
Thus, if the galaxy has experienced zero mergers and the gas content is yet close to zero, the spin magnitude can still be close to  maximum, as no event intervened to change it.
As galaxy mergers can trigger strong quasar activity~\citep{dimatteoetal05}, gas can be efficiently blown away from the central BH activity, and thus there is strong probability merging galaxies end up into gas-poor remnants.

\section{Conclusions and Discussion}
\label{section:conclusion}

Using hydrodynamical cosmological simulations with BH growth and AGN feedback, we estimate the channels through which BHs gain most of their mass, BH-BH coalescences or gas accretion, and we post-processed the evolution of BH spins with analytical prescriptions for changing spins through accretion of gas and BH coalescence.
Our findings are the following:

\begin{itemize}
\item{Low mass BHs, $M_{\rm BH}\lesssim10^8\, \rm M_\odot$, have gained most of their mass through direct accretion of gas, while the most massive BHs are dominated by mergers, $f_{\rm merge}>0.5$ for BHs with $M_{\rm BH}>10^9\, \rm M_\odot$. The more massive a BH is, the earlier BH mergers start to contribute to its mass growth significantly. Most of the BHs that gain 50\% or more of their mass through coalescences have low gas fractions.}
\item{At high redshift, $z>2$,  BHs regularly grow by gas accretion. Except at highest redshifts, the gas possesses a high degree of AM coherency, with small variations of the degree of alignment with the galaxy total angular momentum. At low redshift, $z<2$,  the most massive galaxies have consumed most of their gas. In general, the mean accretion rates are low and AM is random. Gas accretion, in this phase, does not contribute significantly to the BH evolution. Less massive galaxies contain in permanence a reservoir of cold star-forming gas supplying accretion for its central BH (and for star formation).}
\item{Coherent gas accretion is the main driver of BH growth for midsize mass BHs $10^6 \,Ê\rm M_\odot < M_{\rm BH}<10^8 \,Ê\rm M_\odot$. These BHs grow mostly from smoothly capturing gas from the disc  of their host halo. Contribution from binary coalescences to the final BH mass becomes significant for the most massive BHs, $M_{\rm BH}>10^8 \,Ê\rm M_\odot$. As a consequence, midsize mass BHs have spins close to maximal, while the most massive BHs have moderate spin values $|a|\sim0.7$.}
\item{The ratio of accretion rate over Eddington is continuously declining over cosmic time as a results of gas rarefaction in galaxies, while the contribution from mergers is increasing. At high redshift, even the most massive BHs have spin close to their maximum because the influence of mergers is transitory as gas accretion is kept at high rates and maintains spins aligned with their gas surroundings. As a consequence, the radiative efficiency of the accretion process onto BHs is $\epsilon_{\rm r}=0.32$ at high redshift, and lower at low redshift and for the most massive BHs with $\epsilon_{\rm r}=0.1$ for BHs with $M_{\rm BH}>10^9 \,Ê\rm M_\odot$ at $z=0$.}
\item{AGN feedback affects spin evolution indirectly, by decreasing the amount of gas available for BH accretion and its temperature, but not directly by homogenizing the nuclear gas angular momentum.}
\item{Spin direction does not perfectly correlate with the direction of the stellar angular momentum of its host galaxy, particularly for the most massive galaxies and gas-poor galaxies. BHs in these galaxies accrete, at very low levels, from the turbulent hot intra-cluster gas that bears no connection to the angular momentum of the stellar component of the galaxies.}
\item{10-30\% of BHs have spins counter-aligned with their host galaxy (``negative spins"). This occurs in galaxies where the central regions have been deprived of cold gas ($M_{\rm BH}\lesssim 10^8 \, \rm M_\odot$) or because BH coalescences have flipped the spin direction ($M_{\rm BH}\gtrsim 10^8 \, \rm M_\odot$). }
\end{itemize}

We have made several assumptions in our model, one of which is that the gas angular momentum evaluated at kpc scales in our simulations maintains its properties, namely its angular momentum direction, down to the sub-parsec scales close to the BH.
We discussed the robustness of this assumption in Appendix~\ref{section:zoom-halo}, using a 10 pc resolution high-redshift simulation and an isolated halo (and in greater details in paper II), and find that the gas angular momentum within the bulge of the galaxy shows a significant degree of correlation with the gas angular momentum on  galactic scales. Simulations by~\cite{hopkinsetal12} do not find such strong correlations in merging or bar-unstable galaxies, whereas simulations by~\cite{dottietal10} and~\cite{maioetal13} suggest that BHs are able to quickly align with their gas surroundings without varying much in direction when embedded in dense circumnuclear gas discs.

We investigated the effect of adding some degree of incoherence to the accreted gas angular momentum in Appendix~\ref{section:incoherence}.
Even a small amount of coherence \citep[gas is accreted in the semi-sphere defined by the kpc-scale  gas angular momentum, consistent with the degree of misalignment suggested by][]{hopkinsetal12} is sufficient to get spin values consistent with those obtained for purely coherent gas accretion.
On the other hand, if the accretion process is treated as purely random, the lowest mass BHs have the lowest spin values and the highest mass BHs have the largest spin values because in this conditions binary coalescences tend to spin-up BHs \citep{Berti2008}.

In our model we do not treat BH growth and AGN feedback consistently with the spin evolution.
The value of the spin in fact interferes with the Eddington limit of luminosity that sets a maximum accretion rate on the BH.
If luminosity is capped at the Eddington limit, i.e., $\chi=1$, the accretion rate on the BH scales as $\dot{M}=4 \pi G M_{\rm BH} m_{\rm p}/(\epsilon_{\rm r} \sigma_{\rm T} c)$. 
In our simulations, we set $\epsilon_{\rm r}=0.1$, corresponding to a BH spin $a\sim 0.7$, but for a spin $a=0.998$ $\epsilon_{\rm r}=0.32$ and the BH growth is slower by a factor three \citep{King2006}. 
In turn, this affects also the AGN energy release through feedback (see section 2.3) and the directions of jets. 
Therefore, there is room for potential improvement of the method by taking these effects into account directly into the code (which is done on-the-fly in paper I and II), rather than in post-processing.
We note that the efficiency of AGN feedback is of lower importance as models of AGN feedback are already calibrated with lower efficiencies to account for our inability to capture for the detailed small-scale physics.

The direction of radio jets depends on the orientation of BH spins as jets emerge perpendicular to the inner part of the accretion disc which aligns with the spin.
The capability of jets to change directions is possibly a crucial phenomenon to help self-regulate the cooling flows in the centre of clusters of galaxies~\citep{babuletal13}, even though explosive jet outbursts could isotropise the energy output by shocks, sound waves or turbulence~\citep{sandersetal07, duboisetal10, gasparietal11}. Therefore models of jet formation in galaxy clusters could be improved by using the direction of spins rather than local gas angular momentum. 
As we have seen in this paper, gas angular momentum in the centre of galaxy clusters rapidly changes its direction because of the high level of turbulence in the hot intra-cluster gas, whereas the BH spin can keep its direction for several hundreds of million years. 
This self-consistent implementation will be the subject of forthcoming work.

\section*{Acknowledgments}
YD thanks Christophe Pichon for stimulating discussions and helpful comments, Taysun Kimm for his helpful comments, Julien Devriendt and Adrianne Slyz for their continuous support and providing access to Oxford computational resources.
MV thanks Margo Aller for helpful discussions and enlightening suggestions. 
The simulations presented here were run on the DiRAC facility jointly funded by STFC, the Large Facilities Capital Fund of BIS and the University of Oxford. 
This research is part of the Horizon-UK project. 
YD and JS acknowledge support by the ERC advanced grant (Dark Matters). MV acknowledges funding support from NASA, through award ATP NNX10AC84G, and from a Marie Curie Career Integration grant (PCIG10-GA-2011-303609). 
YD thanks Jonathan Patterson and St\'ephane Rouberol for their technical support during the course of this work.

\bibliographystyle{mn2e}
\bibliography{author}

\begin{thebibliography}{}

\bibitem[\protect\citeauthoryear{{Aubert}, {Pichon} \& {Colombi}}{{Aubert}
  et~al.}{2004}]{aubertetal04}
{Aubert} D.,  {Pichon} C.,    {Colombi} S.,  2004, \mnras, 352, 376

\bibitem[\protect\citeauthoryear{{Babul}, {Sharma} \& {Reynolds}}{{Babul}
  et~al.}{2012}]{babuletal13}
{Babul} A.,  {Sharma} P.,    {Reynolds} C.~S.,  2012, ArXiv e-prints

\bibitem[\protect\citeauthoryear{{Barausse}}{{Barausse}}{2012}]{barausse12}
{Barausse} E.,  2012, \mnras, 423, 2533

\bibitem[\protect\citeauthoryear{{Bardeen}}{{Bardeen}}{1970}]{bardeen70}
{Bardeen} J.~M.,  1970, \nat, 226, 64

\bibitem[\protect\citeauthoryear{{Battye} \& {Browne}}{{Battye} \&
  {Browne}}{2009}]{BattyeBrowne2009}
{Battye} R.~A.,  {Browne} I.~W.~A.,  2009, \mnras, 399, 1888

\bibitem[\protect\citeauthoryear{{Bellovary}, {Brooks}, {Volonteri},
  {Governato}, {Quinn} \& {Wadsley}}{{Bellovary}
  et~al.}{2013}]{bellovaryetal13}
{Bellovary} J.,  {Brooks} A.,  {Volonteri} M.,  {Governato} F.,  {Quinn} T.,
  {Wadsley} J.,  2013, \apj, 779, 136

\bibitem[\protect\citeauthoryear{{Berti} \& {Volonteri}}{{Berti} \&
  {Volonteri}}{2008}]{Berti2008}
{Berti} E.,  {Volonteri} M.,  2008, \apj, 684, 822

\bibitem[\protect\citeauthoryear{{Bondi}}{{Bondi}}{1952}]{bondi52}
{Bondi} H.,  1952, \mnras, 112, 195

\bibitem[\protect\citeauthoryear{{Booth} \& {Schaye}}{{Booth} \&
  {Schaye}}{2009}]{booth&schaye09}
{Booth} C.~M.,  {Schaye} J.,  2009, \mnras, 398, 53

\bibitem[\protect\citeauthoryear{{Bournaud}, {Dekel}, {Teyssier}, {Cacciato},
  {Daddi}, {Juneau} \& {Shankar}}{{Bournaud} et~al.}{2011}]{bournaudetal11}
{Bournaud} F.,  {Dekel} A.,  {Teyssier} R.,  {Cacciato} M.,  {Daddi} E.,
  {Juneau} S.,    {Shankar} F.,  2011, \apjl, 741, L33

\bibitem[\protect\citeauthoryear{{Bower}, {Benson}, {Malbon}, {Helly}, {Frenk},
  {Baugh}, {Cole} \& {Lacey}}{{Bower} et~al.}{2006}]{boweretal06}
{Bower} R.~G.,  {Benson} A.~J.,  {Malbon} R.,  {Helly} J.~C.,  {Frenk} C.~S.,
  {Baugh} C.~M.,  {Cole} S.,    {Lacey} C.~G.,  2006, \mnras, 370, 645

\bibitem[\protect\citeauthoryear{{Brenneman} et~al.,}{{Brenneman}
  et~al.}{2011}]{brennemanetal11}
{Brenneman} L.~W.  et~al., 2011, \apj, 736, 103

\bibitem[\protect\citeauthoryear{{Browne} \& {Battye}}{{Browne} \&
  {Battye}}{2010}]{BrowneBattye2010}
{Browne} I.~W.~A.,  {Battye} R.~A.,  2010, in {Maraschi} L.,  {Ghisellini} G.,
  {Della Ceca} R.,   {Tavecchio} F.,  eds,  Astronomical Society of the Pacific
  Conference Series Vol. 427, Accretion and Ejection in AGN: a Global View.
  p.~365

\bibitem[\protect\citeauthoryear{{Chapon}, {Mayer} \& {Teyssier}}{{Chapon}
  et~al.}{2013}]{chaponetal13}
{Chapon} D.,  {Mayer} L.,    {Teyssier} R.,  2013, \mnras, 429, 3114

\bibitem[\protect\citeauthoryear{{Croton} et~al.,}{{Croton}
  et~al.}{2006}]{crotonetal06}
{Croton} D.~J.  et~al., 2006, \mnras, 365, 11

\bibitem[\protect\citeauthoryear{{Daly}}{{Daly}}{2009}]{2009ApJ...696L..32D}
{Daly} R.~A.,  2009, \apjl, 696, L32

\bibitem[\protect\citeauthoryear{{Daly}}{{Daly}}{2011}]{2011MNRAS.414.1253D}
{Daly} R.~A.,  2011, \mnras, 414, 1253

\bibitem[\protect\citeauthoryear{{Daly} \& {Sprinkle}}{{Daly} \&
  {Sprinkle}}{2013}]{2013arXiv1312.4862D}
{Daly} R.~A.,  {Sprinkle} T.~B.,  2013, ArXiv e-prints

\bibitem[\protect\citeauthoryear{{Di Matteo}, {Khandai}, {DeGraf}, {Feng},
  {Croft}, {Lopez} \& {Springel}}{{Di Matteo} et~al.}{2012}]{dimatteoetal12}
{Di Matteo} T.,  {Khandai} N.,  {DeGraf} C.,  {Feng} Y.,  {Croft} R.~A.~C.,
  {Lopez} J.,    {Springel} V.,  2012, \apjl, 745, L29

\bibitem[\protect\citeauthoryear{{Di Matteo}, {Springel} \& {Hernquist}}{{Di
  Matteo} et~al.}{2005}]{dimatteoetal05}
{Di Matteo} T.,  {Springel} V.,    {Hernquist} L.,  2005, \nat, 433, 604

\bibitem[\protect\citeauthoryear{{Dotti}, {Colpi}, {Pallini}, {Perego} \&
  {Volonteri}}{{Dotti} et~al.}{2013}]{dottietal13}
{Dotti} M.,  {Colpi} M.,  {Pallini} S.,  {Perego} A.,    {Volonteri} M.,  2013,
  \apj, 762, 68

\bibitem[\protect\citeauthoryear{{Dotti}, {Volonteri}, {Perego}, {Colpi},
  {Ruszkowski} \& {Haardt}}{{Dotti} et~al.}{2010}]{dottietal10}
{Dotti} M.,  {Volonteri} M.,  {Perego} A.,  {Colpi} M.,  {Ruszkowski} M.,
  {Haardt} F.,  2010, \mnras, 402, 682

\bibitem[\protect\citeauthoryear{{Dubois}, {Devriendt}, {Slyz} \&
  {Teyssier}}{{Dubois} et~al.}{2010}]{duboisetal10}
{Dubois} Y.,  {Devriendt} J.,  {Slyz} A.,    {Teyssier} R.,  2010, \mnras, 409,
  985

\bibitem[\protect\citeauthoryear{{Dubois}, {Devriendt}, {Slyz} \&
  {Teyssier}}{{Dubois} et~al.}{2012}]{duboisetal12agnmodel}
{Dubois} Y.,  {Devriendt} J.,  {Slyz} A.,    {Teyssier} R.,  2012, \mnras, 420,
  2662

\bibitem[\protect\citeauthoryear{{Dubois}, {Pichon}, {Devriendt}, {Silk},
  {Haehnelt}, {Kimm} \& {Slyz}}{{Dubois} et~al.}{2013}]{duboisetal13}
{Dubois} Y.,  {Pichon} C.,  {Devriendt} J.,  {Silk} J.,  {Haehnelt} M.,  {Kimm}
  T.,    {Slyz} A.,  2013, \mnras, 428, 2885

\bibitem[\protect\citeauthoryear{{Dubois}, {Pichon}, {Haehnelt}, {Kimm},
  {Slyz}, {Devriendt} \& {Pogosyan}}{{Dubois}
  et~al.}{2012}]{duboisetal12angmom}
{Dubois} Y.,  {Pichon} C.,  {Haehnelt} M.,  {Kimm} T.,  {Slyz} A.,  {Devriendt}
  J.,    {Pogosyan} D.,  2012, \mnras, 423, 3616

\bibitem[\protect\citeauthoryear{{Dubois} \& {Teyssier}}{{Dubois} \&
  {Teyssier}}{2008}]{dubois&teyssier08winds}
{Dubois} Y.,  {Teyssier} R.,  2008, \aap, 477, 79

\bibitem[\protect\citeauthoryear{{Dubois}, {Volonteri}, {Silk}, {Devriendt} \&
  {Slyz}}{{Dubois} et~al.}{2014}]{duboisetalpaper3}
{Dubois} Y.,  {Volonteri} M.,  {Silk} J.,  {Devriendt} J.,    {Slyz} A.,  2014,
  ArXiv e-prints

\bibitem[\protect\citeauthoryear{{Fanidakis}, {Baugh}, {Benson}, {Bower},
  {Cole}, {Done} \& {Frenk}}{{Fanidakis} et~al.}{2011}]{fanidakisetal11}
{Fanidakis} N.,  {Baugh} C.~M.,  {Benson} A.~J.,  {Bower} R.~G.,  {Cole} S.,
  {Done} C.,    {Frenk} C.~S.,  2011, \mnras, 410, 53

\bibitem[\protect\citeauthoryear{{Feng}, {Di Matteo}, {Croft} \&
  {Khandai}}{{Feng} et~al.}{2013}]{fengetal13}
{Feng} Y.,  {Di Matteo} T.,  {Croft} R.,    {Khandai} N.,  2013, ArXiv e-prints

\bibitem[\protect\citeauthoryear{{Gallimore}, {Axon}, {O'Dea}, {Baum} \&
  {Pedlar}}{{Gallimore} et~al.}{2006}]{Gallimore2006}
{Gallimore} J.~F.,  {Axon} D.~J.,  {O'Dea} C.~P.,  {Baum} S.~A.,    {Pedlar}
  A.,  2006, \aj, 132, 546

\bibitem[\protect\citeauthoryear{{Gammie}, {Shapiro} \& {McKinney}}{{Gammie}
  et~al.}{2004}]{gammieetal04}
{Gammie} C.~F.,  {Shapiro} S.~L.,    {McKinney} J.~C.,  2004, \apj, 602, 312

\bibitem[\protect\citeauthoryear{{Gaspari}, {Melioli}, {Brighenti} \&
  {D'Ercole}}{{Gaspari} et~al.}{2011}]{gasparietal11}
{Gaspari} M.,  {Melioli} C.,  {Brighenti} F.,    {D'Ercole} A.,  2011, \mnras,
  411, 349

\bibitem[\protect\citeauthoryear{{Ghisellini}, {Haardt}, {Della Ceca},
  {Volonteri} \& {Sbarrato}}{{Ghisellini} et~al.}{2013}]{Ghisellini2013}
{Ghisellini} G.,  {Haardt} F.,  {Della Ceca} R.,  {Volonteri} M.,    {Sbarrato}
  T.,  2013, ArXiv e-prints

\bibitem[\protect\citeauthoryear{{Goodman} \& {Tan}}{{Goodman} \&
  {Tan}}{2004}]{goodman&tan04}
{Goodman} J.,  {Tan} J.~C.,  2004, \apj, 608, 108

\bibitem[\protect\citeauthoryear{{Greggio} \& {Renzini}}{{Greggio} \&
  {Renzini}}{1983}]{greggio&renzini83}
{Greggio} L.,  {Renzini} A.,  1983, \aap, 118, 217

\bibitem[\protect\citeauthoryear{{Haardt} \& {Madau}}{{Haardt} \&
  {Madau}}{1996}]{haardt&madau96}
{Haardt} F.,  {Madau} P.,  1996, \apj, 461, 20

\bibitem[\protect\citeauthoryear{{Hopkins}, {Hernquist}, {Hayward} \&
  {Narayanan}}{{Hopkins} et~al.}{2012}]{hopkinsetal12}
{Hopkins} P.~F.,  {Hernquist} L.,  {Hayward} C.~C.,    {Narayanan} D.,  2012,
  \mnras, 425, 1121

\bibitem[\protect\citeauthoryear{{Kellermann} et~al.,}{{Kellermann}
  et~al.}{2004}]{Kellermann2004}
{Kellermann} K.~I.  et~al., 2004, \apj, 609, 539

\bibitem[\protect\citeauthoryear{{Kennicutt}
  Jr.}{{Kennicutt}}{1998}]{kennicutt98}
{Kennicutt} Jr. R.~C.,  1998, \apj, 498, 541

\bibitem[\protect\citeauthoryear{{King}}{{King}}{2003}]{king03}
{King} A.,  2003, \apjl, 596, L27

\bibitem[\protect\citeauthoryear{{King}, {Lubow}, {Ogilvie} \&
  {Pringle}}{{King} et~al.}{2005}]{kingetal05}
{King} A.~R.,  {Lubow} S.~H.,  {Ogilvie} G.~I.,    {Pringle} J.~E.,  2005,
  \mnras, 363, 49

\bibitem[\protect\citeauthoryear{{King} \& {Pringle}}{{King} \&
  {Pringle}}{2006}]{King2006}
{King} A.~R.,  {Pringle} J.~E.,  2006, MNRAS, 373, L90

\bibitem[\protect\citeauthoryear{{King}, {Pringle} \& {Hofmann}}{{King}
  et~al.}{2008}]{kingetal08}
{King} A.~R.,  {Pringle} J.~E.,    {Hofmann} J.~A.,  2008, \mnras, 385, 1621

\bibitem[\protect\citeauthoryear{{King}, {Pringle} \& {Livio}}{{King}
  et~al.}{2007}]{kingetal07}
{King} A.~R.,  {Pringle} J.~E.,    {Livio} M.,  2007, \mnras, 376, 1740

\bibitem[\protect\citeauthoryear{{Kinney}, {Schmitt}, {Clarke}, {Pringle},
  {Ulvestad} \& {Antonucci}}{{Kinney} et~al.}{2000}]{Kinney2000}
{Kinney} A.~L.,  {Schmitt} H.~R.,  {Clarke} C.~J.,  {Pringle} J.~E.,
  {Ulvestad} J.~S.,    {Antonucci} R.~R.~J.,  2000, \apj, 537, 152

\bibitem[\protect\citeauthoryear{{Kolykhalov} \& {Syunyaev}}{{Kolykhalov} \&
  {Syunyaev}}{1980}]{kolykhalov&sunyaev80}
{Kolykhalov} P.~I.,  {Syunyaev} R.~A.,  1980, Soviet Astronomy Letters, 6, 357

\bibitem[\protect\citeauthoryear{{Komatsu} et~al.,}{{Komatsu}
  et~al.}{2011}]{komatsuetal11}
{Komatsu} E.  et~al., 2011, \apjs, 192, 18

\bibitem[\protect\citeauthoryear{{Koppitz}, {Pollney}, {Reisswig}, {Rezzolla},
  {Thornburg}, {Diener} \& {Schnetter}}{{Koppitz} et~al.}{2007}]{koppitzetal07}
{Koppitz} M.,  {Pollney} D.,  {Reisswig} C.,  {Rezzolla} L.,  {Thornburg} J.,
  {Diener} P.,    {Schnetter} E.,  2007, Physical Review Letters, 99, 041102

\bibitem[\protect\citeauthoryear{{Krumholz} \& {Tan}}{{Krumholz} \&
  {Tan}}{2007}]{krumholz&tan07}
{Krumholz} M.~R.,  {Tan} J.~C.,  2007, \apj, 654, 304

\bibitem[\protect\citeauthoryear{{Kulier}, {Ostriker}, {Natarajan}, {Lackner}
  \& {Cen}}{{Kulier} et~al.}{2013}]{kulieretal13}
{Kulier} A.,  {Ostriker} J.~P.,  {Natarajan} P.,  {Lackner} C.~N.,    {Cen} R.,
   2013, ArXiv e-prints

\bibitem[\protect\citeauthoryear{{Lagos}, {Padilla} \& {Cora}}{{Lagos}
  et~al.}{2009}]{lagosetal09}
{Lagos} C.~D.~P.,  {Padilla} N.~D.,    {Cora} S.~A.,  2009, \mnras, 395, 625

\bibitem[\protect\citeauthoryear{{Leitherer}, {Ortiz Ot{\'a}lvaro}, {Bresolin},
  {Kudritzki}, {Lo Faro}, {Pauldrach}, {Pettini} \& {Rix}}{{Leitherer}
  et~al.}{2010}]{leithereretal10}
{Leitherer} C.,  {Ortiz Ot{\'a}lvaro} P.~A.,  {Bresolin} F.,  {Kudritzki}
  R.-P.,  {Lo Faro} B.,  {Pauldrach} A.~W.~A.,  {Pettini} M.,    {Rix} S.~A.,
  2010, \apjs, 189, 309

\bibitem[\protect\citeauthoryear{{Leitherer} et~al.,}{{Leitherer}
  et~al.}{1999}]{leithereretal99}
{Leitherer} C.  et~al., 1999, \apjs, 123, 3

\bibitem[\protect\citeauthoryear{{Levine}, {Gnedin} \& {Hamilton}}{{Levine}
  et~al.}{2010}]{levineetal10}
{Levine} R.,  {Gnedin} N.~Y.,    {Hamilton} A.~J.~S.,  2010, \apj, 716, 1386

\bibitem[\protect\citeauthoryear{{Li}, {Wang} \& {Ho}}{{Li}
  et~al.}{2012}]{Li2012}
{Li} Y.-R.,  {Wang} J.-M.,    {Ho} L.~C.,  2012, \apj, 749, 187

\bibitem[\protect\citeauthoryear{{Maio}, {Dotti}, {Petkova}, {Perego} \&
  {Volonteri}}{{Maio} et~al.}{2013}]{maioetal13}
{Maio} U.,  {Dotti} M.,  {Petkova} M.,  {Perego} A.,    {Volonteri} M.,  2013,
  \apj, 767, 37

\bibitem[\protect\citeauthoryear{{Malbon}, {Baugh}, {Frenk} \&
  {Lacey}}{{Malbon} et~al.}{2007}]{Malbon2007}
{Malbon} R.~K.,  {Baugh} C.~M.,  {Frenk} C.~S.,    {Lacey} C.~G.,  2007,
  \mnras, 382, 1394

\bibitem[\protect\citeauthoryear{{Mart{\'{\i}}nez-Sansigre} \&
  {Rawlings}}{{Mart{\'{\i}}nez-Sansigre} \&
  {Rawlings}}{2011a}]{2011MNRAS.418L..84M}
{Mart{\'{\i}}nez-Sansigre} A.,  {Rawlings} S.,  2011a, \mnras, 418, L84

\bibitem[\protect\citeauthoryear{{Mart{\'{\i}}nez-Sansigre} \&
  {Rawlings}}{{Mart{\'{\i}}nez-Sansigre} \&
  {Rawlings}}{2011b}]{2011MNRAS.414.1937M}
{Mart{\'{\i}}nez-Sansigre} A.,  {Rawlings} S.,  2011b, \mnras, 414, 1937

\bibitem[\protect\citeauthoryear{{Merloni} \& {Heinz}}{{Merloni} \&
  {Heinz}}{2008}]{merloni&heinz08}
{Merloni} A.,  {Heinz} S.,  2008, \mnras, 388, 1011

\bibitem[\protect\citeauthoryear{{Middelberg} et~al.,}{{Middelberg}
  et~al.}{2004}]{Middelberg2004}
{Middelberg} E.  et~al., 2004, \aap, 417, 925

\bibitem[\protect\citeauthoryear{{Natarajan} \& {Pringle}}{{Natarajan} \&
  {Pringle}}{1998}]{NatarajanPringle1998}
{Natarajan} P.,  {Pringle} J.~E.,  1998, ApJL, 506, L97

\bibitem[\protect\citeauthoryear{{Ogilvie}}{{Ogilvie}}{1999}]{ogilvie99}
{Ogilvie} G.~I.,  1999, \mnras, 304, 557

\bibitem[\protect\citeauthoryear{{Omma}, {Binney}, {Bryan} \& {Slyz}}{{Omma}
  et~al.}{2004}]{ommaetal04}
{Omma} H.,  {Binney} J.,  {Bryan} G.,    {Slyz} A.,  2004, \mnras, 348, 1105

\bibitem[\protect\citeauthoryear{{Orban de Xivry}, {Davies}, {Schartmann},
  {Komossa}, {Marconi}, {Hicks}, {Engel} \& {Tacconi}}{{Orban de Xivry}
  et~al.}{2011}]{Orban2011}
{Orban de Xivry} G.,  {Davies} R.,  {Schartmann} M.,  {Komossa} S.,  {Marconi}
  A.,  {Hicks} E.,  {Engel} H.,    {Tacconi} L.,  2011, \mnras, 417, 2721

\bibitem[\protect\citeauthoryear{{Ostriker}}{{Ostriker}}{1999}]{ostriker99}
{Ostriker} E.~C.,  1999, \apj, 513, 252

\bibitem[\protect\citeauthoryear{{Patrick}, {Reeves}, {Lobban}, {Porquet} \&
  {Markowitz}}{{Patrick} et~al.}{2011}]{patricketal11}
{Patrick} A.~R.,  {Reeves} J.~N.,  {Lobban} A.~P.,  {Porquet} D.,
  {Markowitz} A.~G.,  2011, \mnras, 416, 2725

\bibitem[\protect\citeauthoryear{{Perego}, {Dotti}, {Colpi} \&
  {Volonteri}}{{Perego} et~al.}{2009}]{Perego2009}
{Perego} A.,  {Dotti} M.,  {Colpi} M.,    {Volonteri} M.,  2009, MNRAS, 399,
  2249

\bibitem[\protect\citeauthoryear{{Pringle}}{{Pringle}}{1981}]{pringle81}
{Pringle} J.~E.,  1981, \araa, 19, 137

\bibitem[\protect\citeauthoryear{{Reynolds}}{{Reynolds}}{2013}]{reynoldsetal13}
{Reynolds} C.~S.,  2013, SSR

\bibitem[\protect\citeauthoryear{{Reynolds}, {Brenneman}, {Lohfink}, {Trippe},
  {Miller}, {Fabian} \& {Nowak}}{{Reynolds} et~al.}{2012}]{reynoldsetal12}
{Reynolds} C.~S.,  {Brenneman} L.~W.,  {Lohfink} A.~M.,  {Trippe} M.~L.,
  {Miller} J.~M.,  {Fabian} A.~C.,    {Nowak} M.~A.,  2012, \apj, 755, 88

\bibitem[\protect\citeauthoryear{{Rezzolla}, {Barausse}, {Dorband}, {Pollney},
  {Reisswig}, {Seiler} \& {Husa}}{{Rezzolla} et~al.}{2008}]{rezzollaetal08}
{Rezzolla} L.,  {Barausse} E.,  {Dorband} E.~N.,  {Pollney} D.,  {Reisswig} C.,
   {Seiler} J.,    {Husa} S.,  2008, \prd, 78, 044002

\bibitem[\protect\citeauthoryear{{Risaliti} et~al.,}{{Risaliti}
  et~al.}{2013}]{risalitietal13}
{Risaliti} G.  et~al., 2013, \nat, 494, 449

\bibitem[\protect\citeauthoryear{{Sanders} \& {Fabian}}{{Sanders} \&
  {Fabian}}{2007}]{sandersetal07}
{Sanders} J.~S.,  {Fabian} A.~C.,  2007, \mnras, 381, 1381

\bibitem[\protect\citeauthoryear{{S{\c a}dowski}, {Bursa}, {Abramowicz},
  {Klu{\'z}niak}, {Lasota}, {Moderski} \& {Safarzadeh}}{{S{\c a}dowski}
  et~al.}{2011}]{sadowskietal11}
{S{\c a}dowski} A.,  {Bursa} M.,  {Abramowicz} M.,  {Klu{\'z}niak} W.,
  {Lasota} J.-P.,  {Moderski} R.,    {Safarzadeh} M.,  2011, \aap, 532, A41

\bibitem[\protect\citeauthoryear{{Scheuer} \& {Feiler}}{{Scheuer} \&
  {Feiler}}{1996}]{Scheuer1996}
{Scheuer} P.~A.~G.,  {Feiler} R.,  1996, MNRAS, 282, 291

\bibitem[\protect\citeauthoryear{{Shakura} \& {Sunyaev}}{{Shakura} \&
  {Sunyaev}}{1973}]{shakura&sunyaev73}
{Shakura} N.~I.,  {Sunyaev} R.~A.,  1973, \aap, 24, 337

\bibitem[\protect\citeauthoryear{{Shankar}, {Weinberg} \&
  {Miralda-Escud{\'e}}}{{Shankar} et~al.}{2013}]{Shankar2013}
{Shankar} F.,  {Weinberg} D.~H.,    {Miralda-Escud{\'e}} J.,  2013, \mnras,
  428, 421

\bibitem[\protect\citeauthoryear{{Shapiro}}{{Shapiro}}{2005}]{shapiro2005}
{Shapiro} S.~L.,  2005, \apj, 620, 59

\bibitem[\protect\citeauthoryear{{Sijacki}, {Springel}, {Di Matteo} \&
  {Hernquist}}{{Sijacki} et~al.}{2007}]{sijackietal07}
{Sijacki} D.,  {Springel} V.,  {Di Matteo} T.,    {Hernquist} L.,  2007,
  \mnras, 380, 877

\bibitem[\protect\citeauthoryear{{Sijacki}, {Springel} \& {Haehnelt}}{{Sijacki}
  et~al.}{2009}]{sijackietal09}
{Sijacki} D.,  {Springel} V.,    {Haehnelt} M.~G.,  2009, \mnras, 400, 100

\bibitem[\protect\citeauthoryear{{Silk} \& {Rees}}{{Silk} \&
  {Rees}}{1998}]{silk&rees98}
{Silk} J.,  {Rees} M.~J.,  1998, \aap, 331, L1

\bibitem[\protect\citeauthoryear{{Sutherland} \& {Dopita}}{{Sutherland} \&
  {Dopita}}{1993}]{sutherland&dopita93}
{Sutherland} R.~S.,  {Dopita} M.~A.,  1993, \apjs, 88, 253

\bibitem[\protect\citeauthoryear{{Teyssier}}{{Teyssier}}{2002}]{teyssier02}
{Teyssier} R.,  2002, \aap, 385, 337

\bibitem[\protect\citeauthoryear{{Teyssier}, {Moore}, {Martizzi}, {Dubois} \&
  {Mayer}}{{Teyssier} et~al.}{2011}]{teyssieretal11}
{Teyssier} R.,  {Moore} B.,  {Martizzi} D.,  {Dubois} Y.,    {Mayer} L.,  2011,
  \mnras, 414, 195

\bibitem[\protect\citeauthoryear{{Thorne}}{{Thorne}}{1974}]{thorne74}
{Thorne} K.~S.,  1974, \apj, 191, 507

\bibitem[\protect\citeauthoryear{{Tweed}, {Devriendt}, {Blaizot}, {Colombi} \&
  {Slyz}}{{Tweed} et~al.}{2009}]{tweedetal09}
{Tweed} D.,  {Devriendt} J.,  {Blaizot} J.,  {Colombi} S.,    {Slyz} A.,  2009,
  \aap, 506, 647

\bibitem[\protect\citeauthoryear{{Verdoes Kleijn} \& {de Zeeuw}}{{Verdoes
  Kleijn} \& {de Zeeuw}}{2005}]{VerdoesKleijn2005}
{Verdoes Kleijn} G.~A.,  {de Zeeuw} P.~T.,  2005, \aap, 435, 43

\bibitem[\protect\citeauthoryear{{Volonteri} \& {Ciotti}}{{Volonteri} \&
  {Ciotti}}{2012}]{Volonteri2013}
{Volonteri} M.,  {Ciotti} L.,  2012, ArXiv e-prints, 1211.6840

\bibitem[\protect\citeauthoryear{{Volonteri}, {Madau}, {Quataert} \&
  {Rees}}{{Volonteri} et~al.}{2005}]{volonterietal05}
{Volonteri} M.,  {Madau} P.,  {Quataert} E.,    {Rees} M.~J.,  2005, \apj, 620,
  69

\bibitem[\protect\citeauthoryear{{Volonteri}, {Sikora} \& {Lasota}}{{Volonteri}
  et~al.}{2007}]{volonterietal07}
{Volonteri} M.,  {Sikora} M.,    {Lasota} J.-P.,  2007, \apj, 667, 704

\bibitem[\protect\citeauthoryear{{Volonteri}, {Sikora}, {Lasota} \&
  {Merloni}}{{Volonteri} et~al.}{2013}]{volonterietal13}
{Volonteri} M.,  {Sikora} M.,  {Lasota} J.-P.,    {Merloni} A.,  2013, \apj,
  775, 94

\bibitem[\protect\citeauthoryear{{Wyithe} \& {Loeb}}{{Wyithe} \&
  {Loeb}}{2003}]{wyithe&loeb03}
{Wyithe} J.~S.~B.,  {Loeb} A.,  2003, \apj, 595, 614

\bibitem[\protect\citeauthoryear{{Yu} \& {Tremaine}}{{Yu} \&
  {Tremaine}}{2002}]{yutremaine2002}
{Yu} Q.,  {Tremaine} S.,  2002, \mnras, 335, 965

\end{thebibliography}

\appendix

\section{Coherence of angular momentum between pc and kpc scales for a multiphase interstellar medium}
\label{section:zoom-halo}

\begin{figure}
  \centering{\resizebox*{!}{6.2cm}{\includegraphics{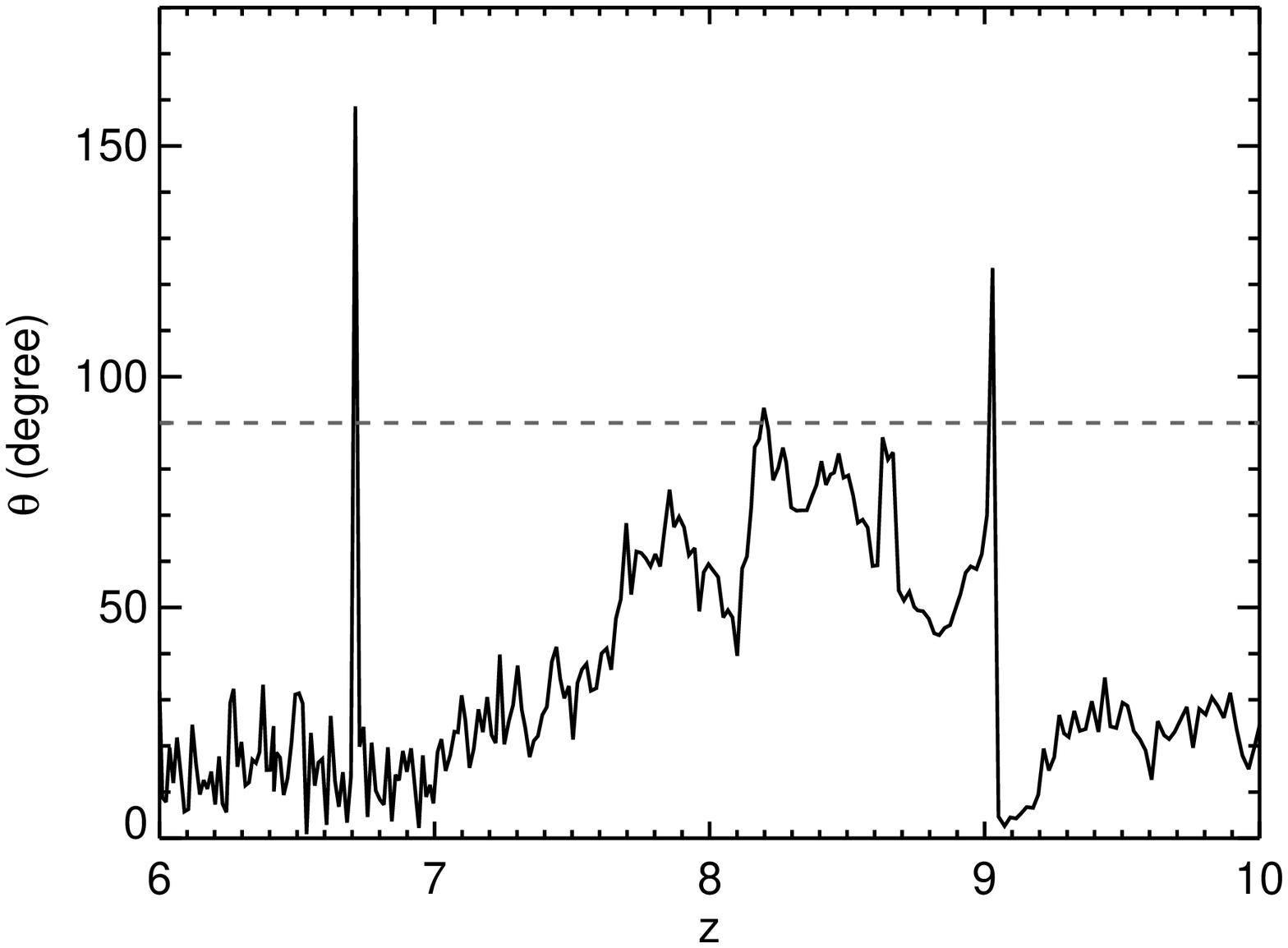}}}
  \caption{Angle between the gas angular momentum of the bulge and the angular momentum of the disc measured at 30 pc and 1 kpc respectively in a zoom cosmological simulation. The dashed horizontal line is 90 degrees above which the gas in the bulge counter-rotates with the gas in the galaxy. }
    \label{fig:angle}
\end{figure}

\begin{figure}
  \centering{\resizebox*{!}{7.2cm}{\includegraphics{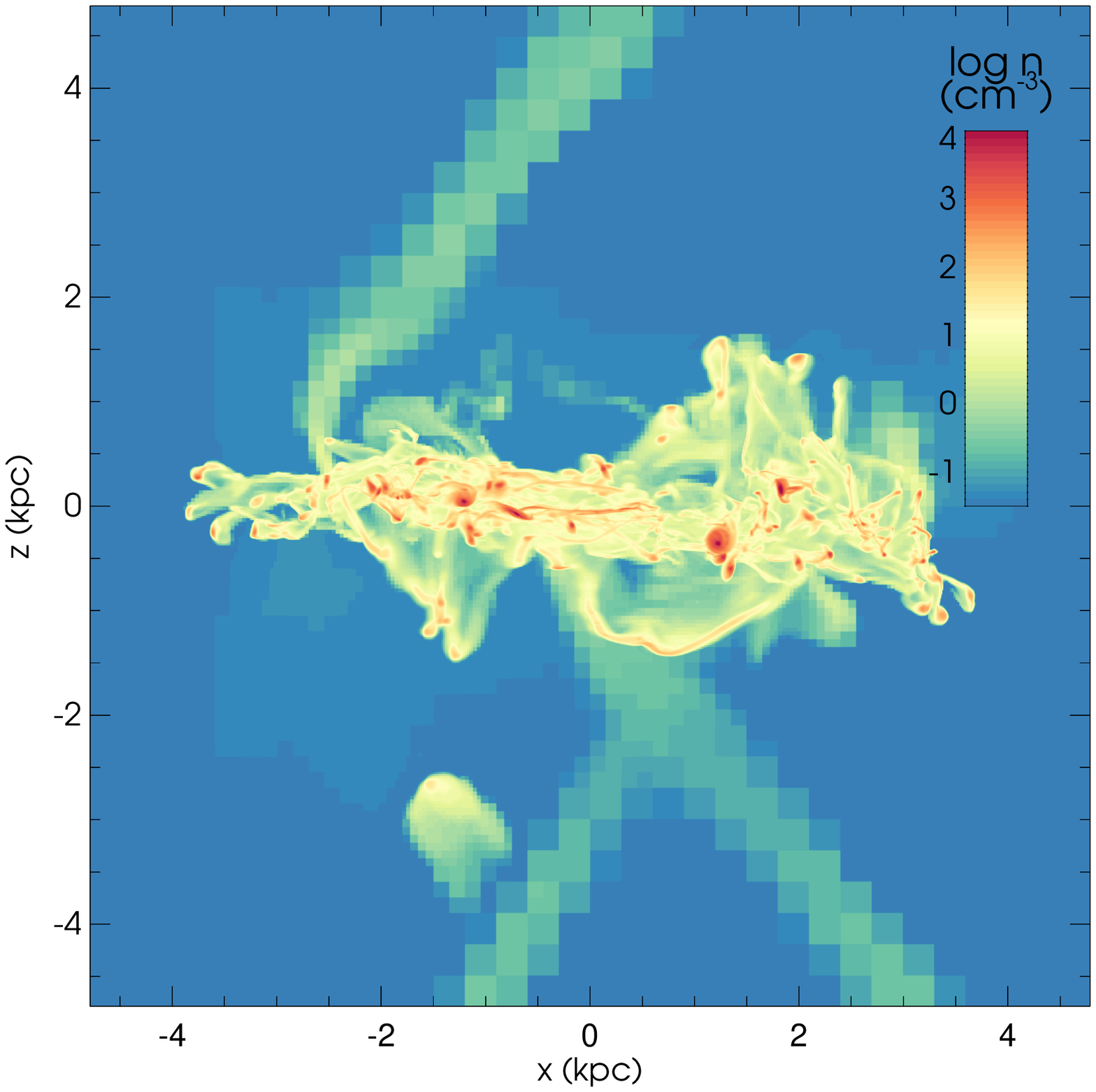}}}\vspace{-0.8cm}
  \centering{\resizebox*{!}{7.2cm}{\includegraphics{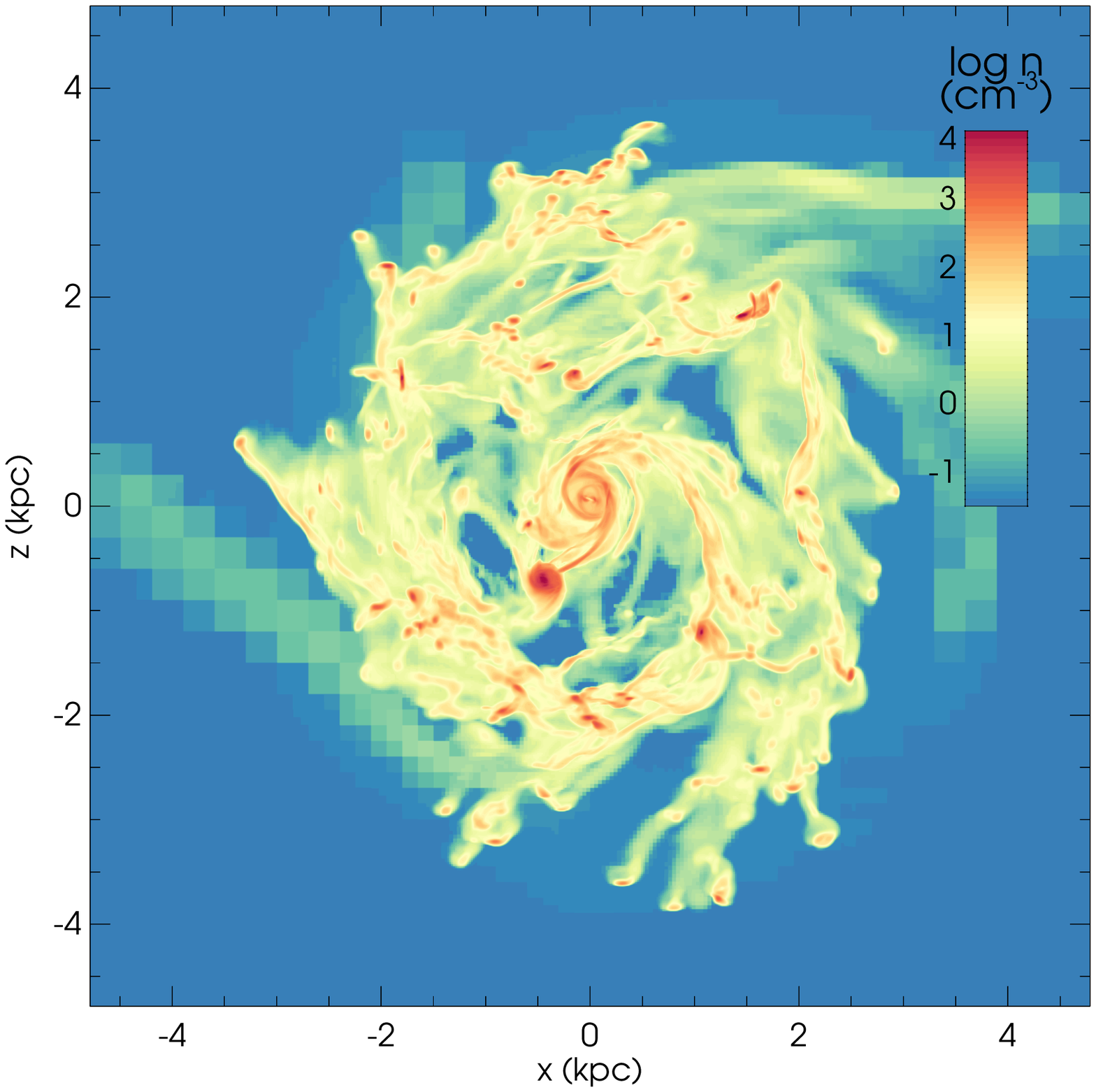}}}
  \caption{Edge-on view (top) and face-on view (bottom) of the gas density in the isolated halo simulation at $t=850 \, \rm Myr$.}
    \label{fig:isolatednfw}
\end{figure}

In order to test and validate the assumption that, on average, the gas angular momentum on kpc scales is an adequate estimate of the gas angular momentum on smaller scales (parsec scales, which are the size of the BH scale of influence), we analysed the zoom-in simulation performed in~\cite{duboisetal13} of a progenitor of a massive cluster at $z=6$ with $\Delta x \simeq 10 \, \rm pc$ resolution with the {\sc ramses} code.
The halo is simulated with the same physics than in the current work, i.e gas cooling, star formation, SN (type II only) and AGN feedback, and its mass is of $5\times 10^{11}\,Ê\rm M_\odot$ at $z=6$.
More detailed information can be found in the paper stated above.

In that simulation, the galaxy is composed of a bulge (50 pc radius) and a disc (1 kpc), which is subject to gas fragmentation due to Toomre instabilities, thus, clumps are visible most of time and are captured by the central bulge due to a rapid migration.
Hence, it is a good case for testing the spatial coherence of gas angular momentum between small scales, and the kpc disc scale.
The procedure is the following: we compute the gas angular momentum within 30 pc and within 1 kpc, both relative to their centre of mass, and we measure the angle $\theta$ between the two vectors.
Fig.~\ref{fig:angle} shows the $\theta$ angle as a function of redshift, and it clearly appears that inner gas close to the bulge has some degree of coherence with the gas on kpc scales.
The abrupt temporal changes in the alignment of both angular momenta at $z=9,8.6, 8.2, 7.9, 6.7$ are due to the capture of a satellite that exerts a large gravitational torque on the central galaxy.
The mergers at $z=9$ and $z=6.7$ cause the rapid counter-rotation of the bulge region relative to the disc, but this effect is not long-lasting (see paper II for details).

\begin{figure}
  \centering{\resizebox*{!}{6.2cm}{\includegraphics{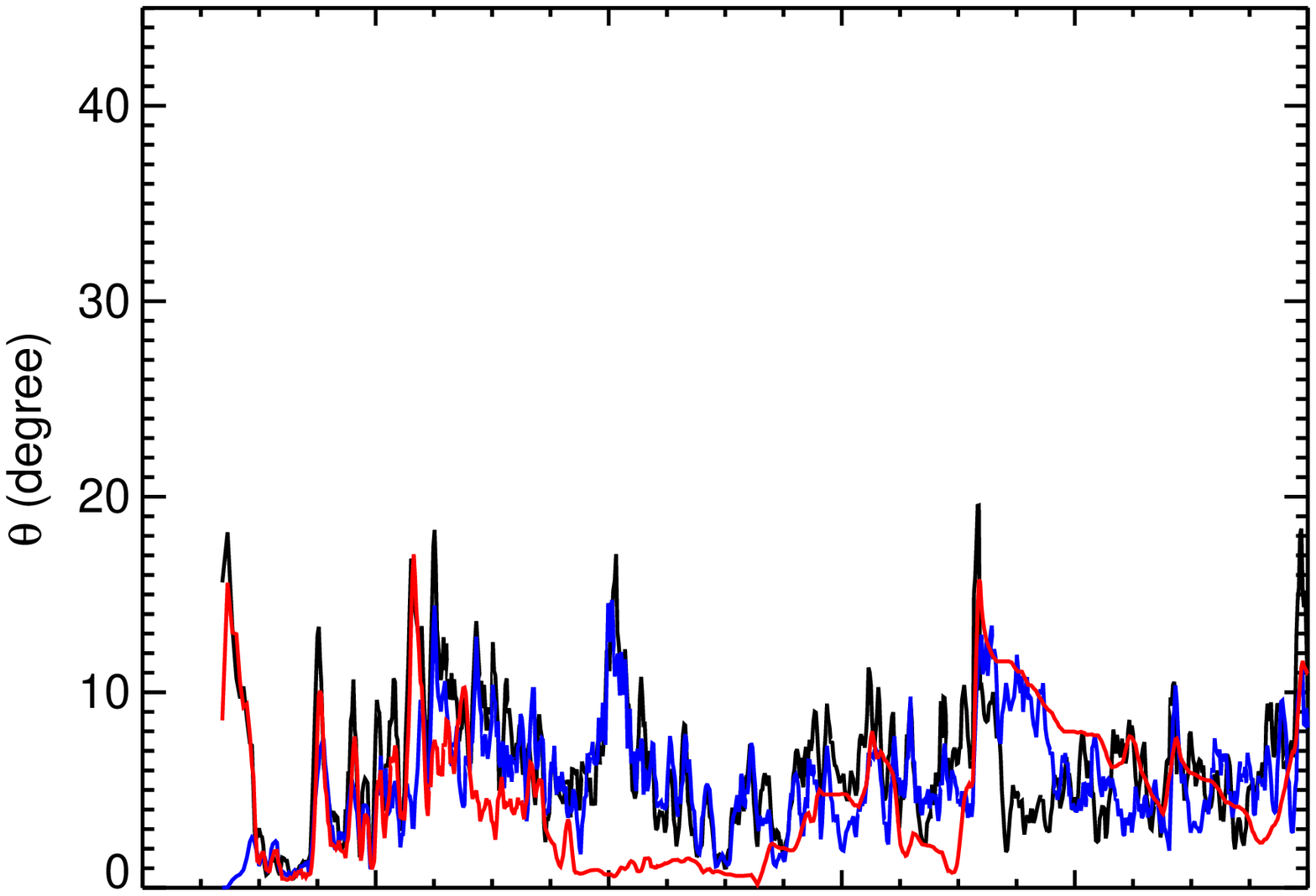}}}\vspace{-1.2cm}
  \centering{\resizebox*{!}{6.2cm}{\includegraphics{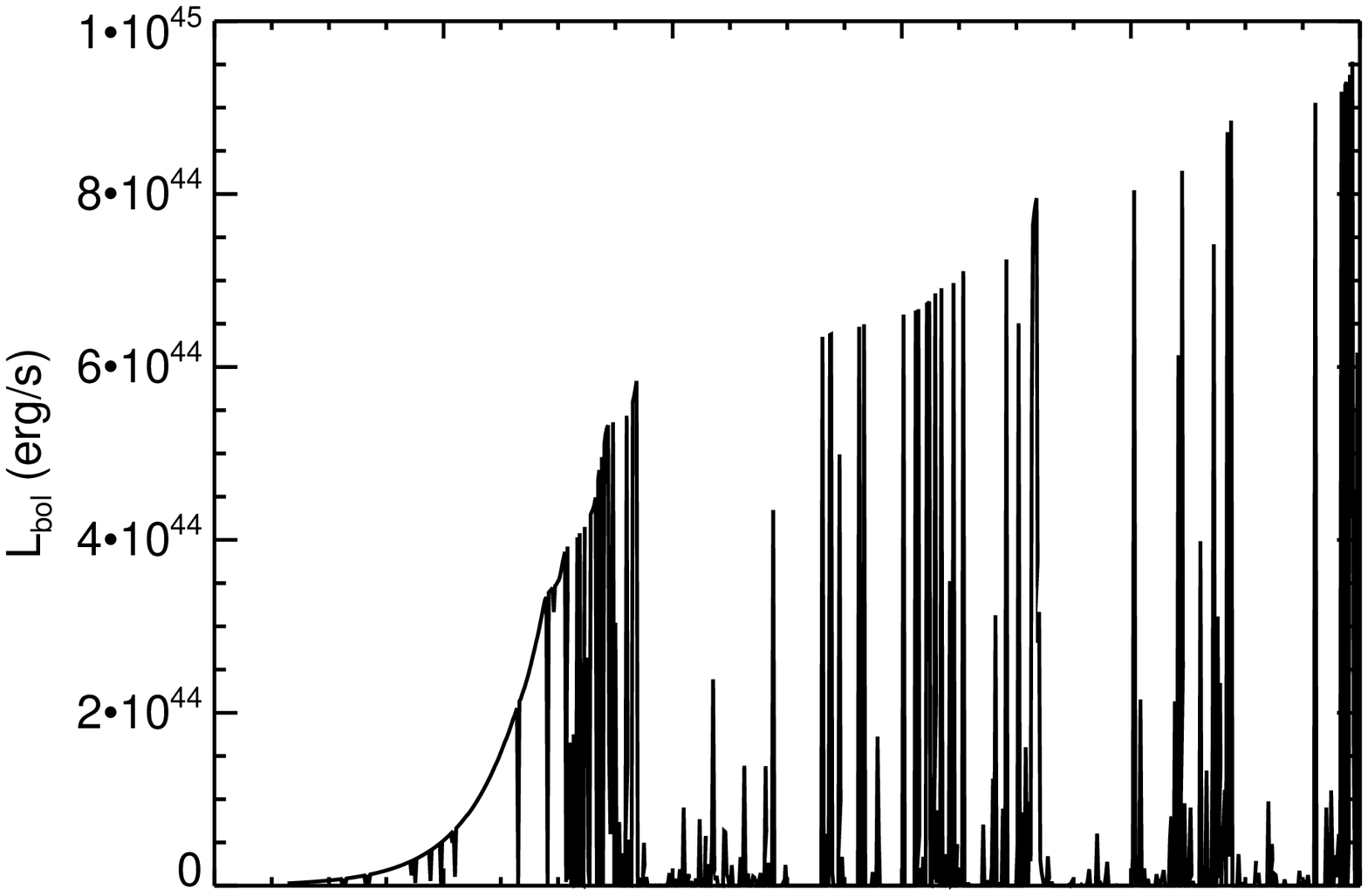}}}\vspace{-1.2cm}
  \centering{\resizebox*{!}{6.2cm}{\includegraphics{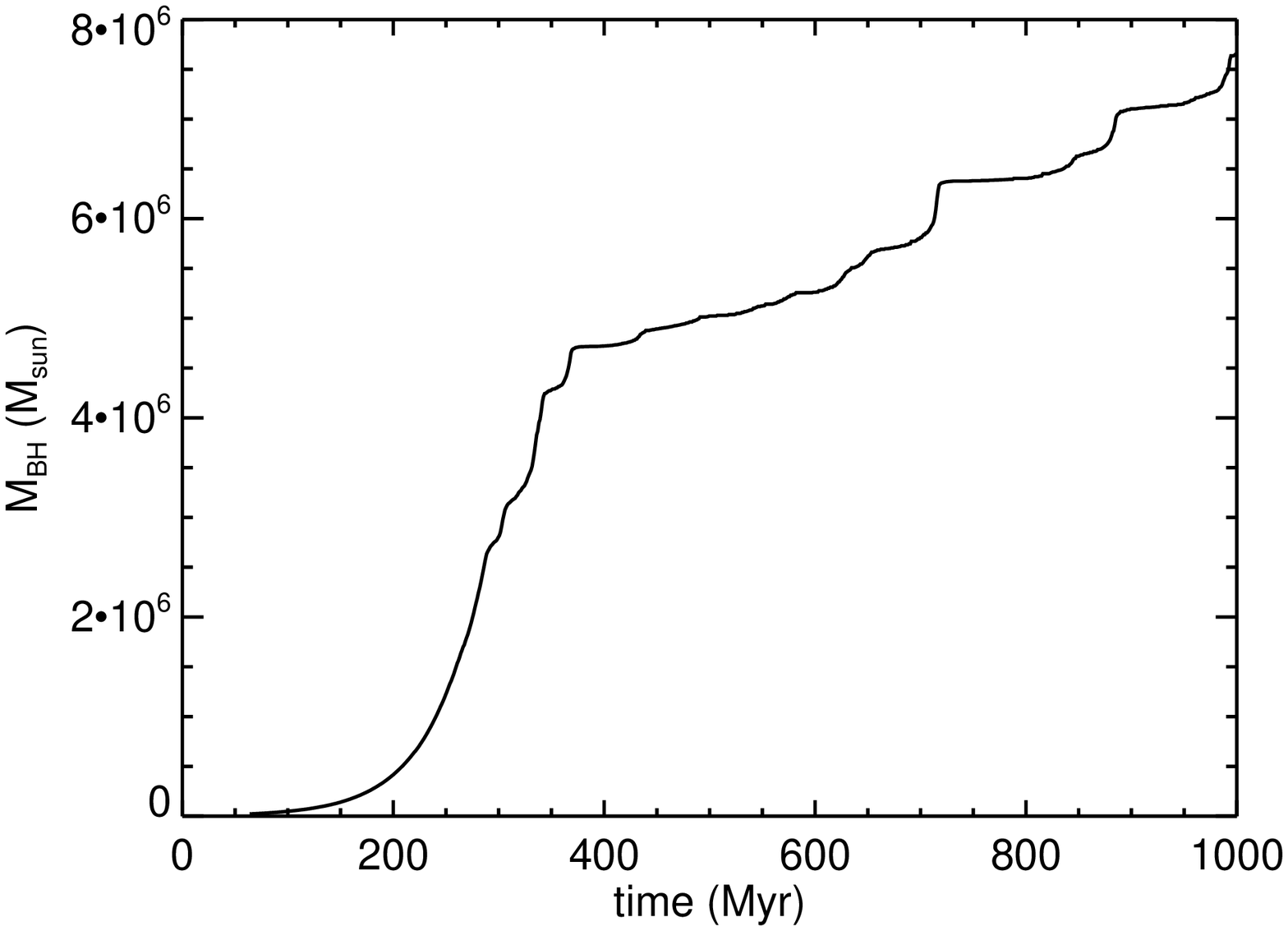}}}
  \caption{\emph{Top:} Inclination angle between the accreted gas angular momentum onto the BH and the gas angular momentum in the galaxy (black), between the accreted gas angular momentum on the BH and the angular momentum in the galaxy (red), and between the BH spin and the accreted gas angular momentum onto the BH (blue); \emph{middle:} bolometric luminosity; \emph{bottom:} BH mass as a function of time for an isolated disc galaxy. The BH follows an initial Eddington-limited growth phase followed by multiple accretion of dense clump of gas that triggers strong quasar activity. Even though gas accretion is powered by these compact clumps of gas at late times, their angular momentum show a high degree of alignment with the BH spin and with the global disc angular momentum.}
    \label{fig:mbhvstime}
\end{figure}

We extended this test in convergence by setting up an isolated halo simulation of a $M_{\rm vir}=10^{12}\, \rm M_\odot$ with NFW profile that has a concentration of $c=3.5$, spin of $\lambda=0.04$, and a gas fraction of $f_{\rm gas}=0.15$. 
Gas is initially in hydrostatic equilibrium in its own potential and those defined by the DM particles composing the dark halo.
DM mass is resolution is $M_{\rm DM, res}=1.7 \times 10^6 \, \rm M_\odot$, and the minimum cell size is of $\Delta x=9\, \rm pc$.
The threshold of star formation is $n_0=250 \, \rm cm^{-3}$ and a minimum temperature of $1000\, \rm K$ provided by cooling through metal lines.
A polytropic equation of state kicks in above this density threshold with polytropic index $\kappa=2$.
Supernovae feedback and AGN feedback are the same as in the simulation described in the body of the paper.
Once the simulation starts, the halo loses its initial pressure support through gas radiative cooling and a disc of gas forms due to the initial rotational support.
A central BH is inserted after 60 Myr with an initial seed BH mass of $2\times 10^4 \, \rm M_\odot$ and reaches a final mass of $7.5\times 10^6 \, \rm M_\odot$ after 1 Gyr.

Such a high resolution provides a multiphase description of the interstellar medium with cold clumps of star-forming gas embedded in a diffuse warm and turbulent medium. 
Fig.~\ref{fig:isolatednfw} shows the galaxy both edge-on and face-on. 
The galaxy disc shows the presence of dense gas clumps, as well as a dynamical thick disc.
The dense clumps  and the disc are rotationally supported, albeit with angles not precisely aligned with plane of the galaxy, and there is a large degree of alignment with the galaxy and, thus with the central BH fed through this gas. 
Fig.~\ref{fig:mbhvstime} highlights the BH growth (bottom panel) and the global coherence between angular momentum on small and large scales, by showing  the angle $\theta$ between the gas angular momentum within 10 pc and within 1 kpc, that remains close to 10$^\circ$ (top panel).
As a consequence, the angular momentum of the clumps when they reach  the central region adds up constructively to the BH spin. 
In this simulation, we followed the spin evolution of the central BH, and we find that the spin is constant and equal to the maximum value at any time with a strong level of alignment with its direct gas surrounding and with gas on the galactic scale.
The BH reaches its maximum spin value after $37.5 \, \rm Myr$ after its formation time (at $t=64\, \rm Myr$ after simulation startup), which correspond to a mass increase of 3 times the initial BH mass during that amount of time when the BH grows at Eddington.
This simulation is also a perfect test-bed for the effect of AGN feedback on the angular momentum distribution of the galaxy gas. 
At time $t\simeq270\, \rm Myr$ a strong outflow develops from the central AGN forming a large-scale galactic wind, with a bolometric luminosity from the central BH of $L_{\rm bol}= \epsilon_{\rm r}\dot M_{\rm BH}c^2\simeq 2\times 10^{44} \, \rm erg\,s^{-1}$. 
Then, the accretion onto the central BH proceeds through the capture of dense clump of gas migrating from the galactic disc down to the bulge~\citep{bournaudetal11, duboisetal12angmom, duboisetal13} triggering powerful quasar activity with luminosity in the range  $L_{\rm bol}=  10^{44}-10^{45} \, \rm erg\,s^{-1}$ (see Fig.~\ref{fig:mbhvstime}).
Despite this strong interaction between the AGN and the galaxy gas, the overall coherence of the gas in the disk, between 10 pc and 1 kpc, is not strongly perturbed.

As a final note, we have resolution limited to $10\, \rm pc$ and it is well possible that  a more resolved turbulent spectrum would change this result, but complementary simulations performed at sub-pc resolution by~\cite{maioetal13} confirm that, when the accreting gas possesses a definite angular momentum being distributed in a disc, local turbulence, clumpy star formation, infall from recycled stellar wind material and stellar feedback do not induce sufficient turbulence to introduce a high degree of incoherence in the gas that accretes on the BH, even though they do not consider feedback from a central AGN source, which we instead include.  In the case of galaxy mergers or moderately bar-unstable discs, instead,~\cite{hopkinsetal12} find that accretion onto the central BH is much more random though they find more alignment than in the pure isotropic case. In their Figure~1, however, most of the galaxies show a strong alignment with the z-axis down to $\sim$ 1 pc with strong departures from coherency when approaching 0.1 pc.  This is close to their resolution limit  (0.1 pc) and several smoothing lengths may in fact be required to sample the nuclear angular momentum properly.
 
\section{BH spins for different degrees of gas incoherence}
\label{section:incoherence}

We tested the effect of adding some level of incoherence in the measured gas angular momentum when updating the BH spin through gas accretion \citep[see][for a detailed discussion on incoherence]{dottietal13}.
At each coarse time step, we change the measured gas angular momentum with a random angle uniformly distributed between 0 and $\theta_{\rm max}$, where $\theta_{\rm max}=\pi /2$ or $\theta_{\rm max}=\pi$.
The new direction of the gas angular momentum is then used to update the direction of the new spin.

We see that when gas accretion is completely chaotic (direction of gas angular momentum is totally random), it leads to the formation of low-mass and midsize-mass BHs $M_{\rm BH}< 10^8\, \rm M_\odot$ with low spin values of $|a|=0.2-0.5$, which is consistent with results of~\cite{fanidakisetal11}.
The value of spins is increasing for more massive BHs as they build a large fraction of their mass through mergers.
When the accreted gas is randomised with some level of coherence relative to the gas angular momentum provided by the simulation \citep[pointing in the same semi-sphere as the gas angular momentum, consistent with the degree of misalignment suggested by][]{hopkinsetal12}, values of spin are closer to the ones predicted in our reference model.
For $M_{\rm BH}< 10^8\, \rm M_\odot$ BH spins are close to unity, $|a|=0.9-0.95$, and more massive BHs have lower spin values. 
Note that, in this case, values of the spin, on average, never differ more than 10 per cent relative to our reference model.

We applied an extra test, where we assume that the gas angular momentum accreted onto the BH is the one provided by the simulation as in the reference model except when strong merger events happen.
We define a strong merger as the coalescence of BHs with mass ratio larger than $1:10$.
For these pairs of BHs, gas is accreted in a chaotic fashion (random orientations) during 50 Myr which is a typical duration for merger-induced starbursts, and the typical e-folding time of a BH accreting at the Eddington rate.
After 50 Myr, if no more mergers happen, gas accretes with the angular momentum measured from the simulation.
The result of that experiment on BH spins is similar to the reference model: values of the spins are similar to within 1 per cent.
We find that mergers are not responsible for the accretion of cold gas onto BHs.
The vast majority of the gas that contributes to BH growth is accreted while galaxies are not merging and are smoothly acquiring gas from their host halo.

\begin{figure}
  \centering{\resizebox*{!}{6.2cm}{\includegraphics{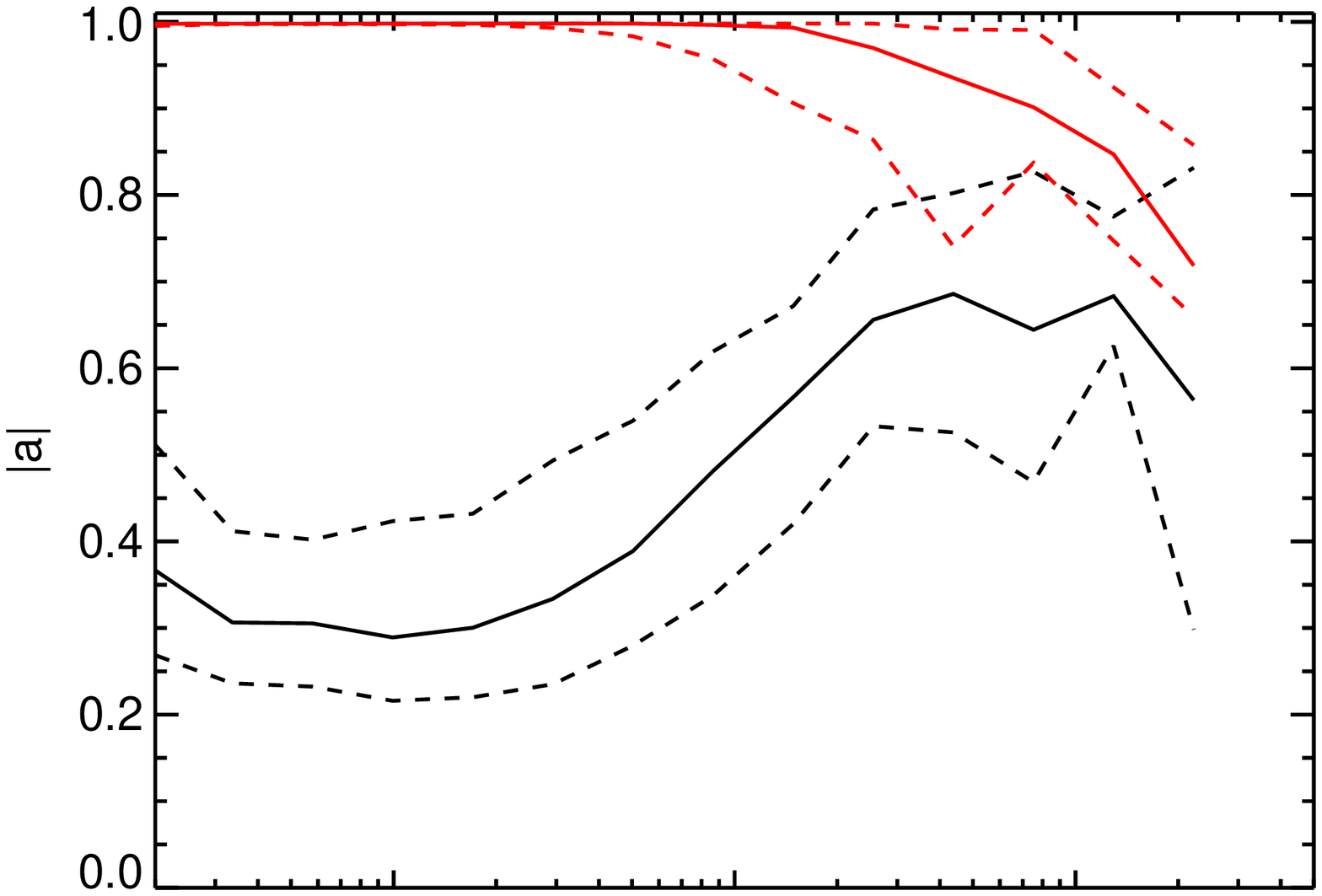}}}\vspace{-1.2cm}
  \centering{\resizebox*{!}{6.2cm}{\includegraphics{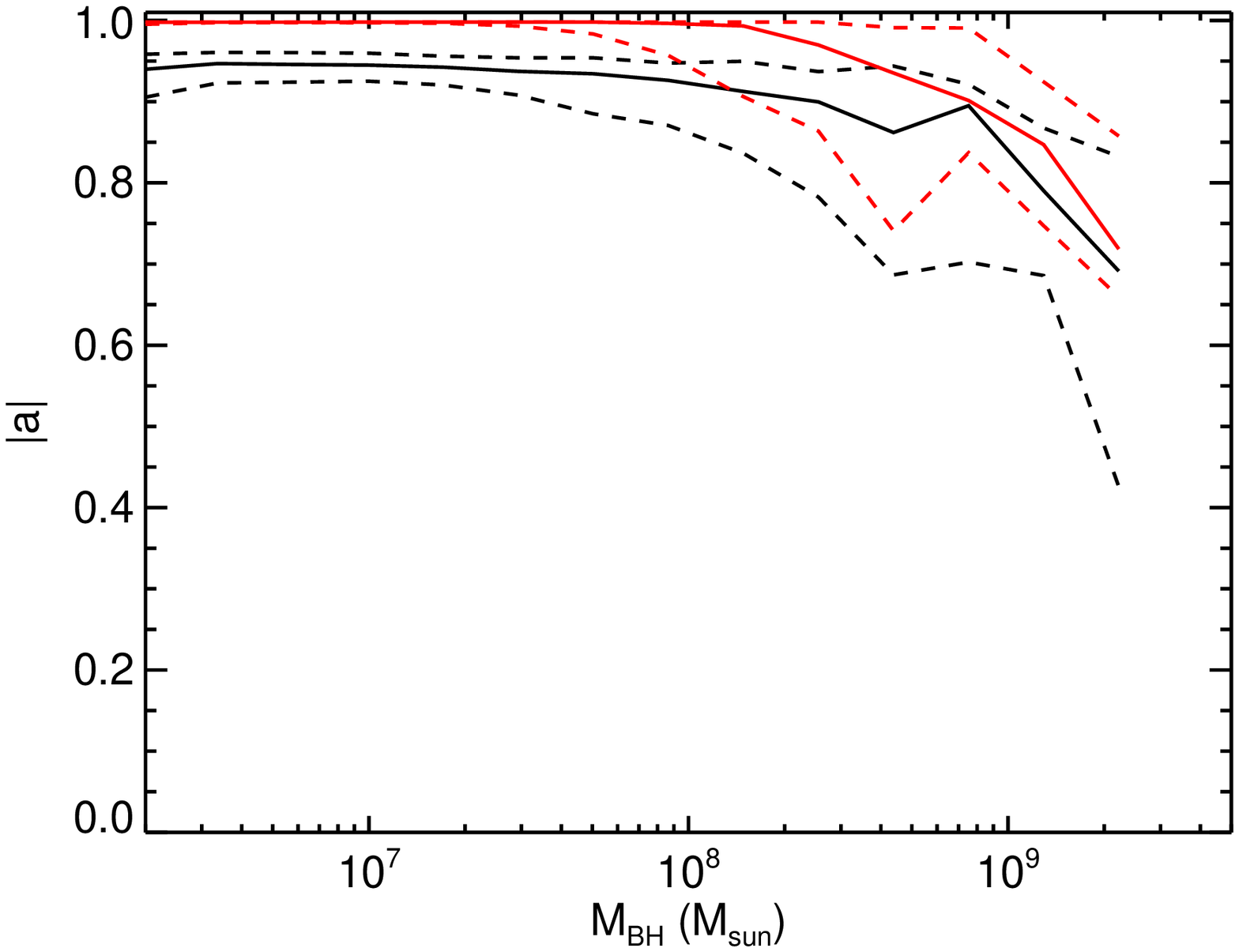}}}
  \caption{Median of the spin as a function of BH mass at $z=0$ (black solid lines) for an angular momentum with an angle respective to the measured gas angular momentum drawn between $[0,\pi]$ (top) and $[0,\pi/2]$ (bottom). Dashed lines show the 20 and 80 percentiles. Red lines corresponds to no random angle (result from Fig~\ref{fig:abhvsmbh_ave}).}
    \label{fig:abhvsmbh_ave_random}
\end{figure}

\end{document}